\documentclass[aps,rmp,onecolumn,groupedaddress,nofootinbib,showkeys,11pt,tightenlines]{revtex4}
\newcommand{\kgo}{(\mbox{\large{$\sqcup\!\!\!\!\sqcap$}}\;+\;m^2+\kappa R)}
\newcommand{\df}{\;\colon\!=}
\newcommand{\fd}{=\colon\;}
\newcommand{\dfff}{\;\colon\!\!\!\!=}

\newcommand{\ul}[1]{\underline{#1}}
\usepackage{amssymb}
\usepackage{amsmath}
\usepackage{amsthm}
\usepackage{graphicx}
\usepackage{cancel}
\usepackage{multido}
\def\R{{\mathbb R}}
\def\N{{\mathbb N}}
\def\C{{\mathbb C}}

\bibpunct{[}{]}{,}{a}{,}{,}

\begin{document}

\pagestyle{myheadings} \markboth{ H. Gottschalk and T. Hack }{ Non-quasifree states on curved spacetimes } \thispagestyle{empty}


\title{On the unitary transformation between non-quasifree and quasifree state spaces and its application to quantum field theory on curved spacetimes}
 \author{Hanno Gottschalk}
 
 \affiliation{Institut f\"ur angewandte Mathematik, Rheinische Friedrich-Wilhelms-Universit\"at Bonn}
 
 \email{gottscha@wiener.iam.uni-bonn.de}
 
 \author{Thomas-Paul Hack}
 
 \affiliation{{I}{I}. Institut f\"ur Theoretische Physik, Universit\"at Hamburg}
 
 \email{thomas-paul.hack@desy.de}
 
 \date{\today}
 
 \begin{abstract}
Using $\star$-calculus on the dual of the Borchers-Uhlmann algebra endowed with a combinatorial co-product, we develop a
method to calculate a unitary transformation relating the GNS representations of a non-quasifree and a quasifree state of the free hermitian scalar field. The motivation for such an analysis and a further result is the fact that a unitary transformation of this kind arises naturally in scattering theory on non-stationary backgrounds. Indeed, employing the perturbation theory of the Yang-Feldman equations with a free CCR field in a quasifree state as an initial condition and making use of extended Feynman graphs, we are able to calculate the Wightman functions of the interacting and outgoing fields in a $\phi^p$-theory on arbitrary curved spacetimes. A further examination then reveals two major features of the aforementioned theory: firstly, the interacting Wightman functions fulfil the basic axioms of hermiticity, invariance, spectrality (on stationary spacetimes), perturbative positivity, and locality. Secondly, the outgoing field is free and fulfils the CCR, but is in general not in a quasifree state in the case of a non-stationary spacetime. In order to obtain a sensible particle picture for the outgoing field and, hence, a description of the scattering process in terms of particles (in asymptotically flat spacetimes), it is thus necessary to compute a unitary transformation of the abovementioned type.
\end{abstract}

\keywords{Interacting quantum fields on curved
spacetimes, Wightman functions, perturbation theory, non-quasifree
states\\[11pt]MSC (2000): 81T20}

 \maketitle

\section {Introduction}
In the theory of quantum fields on curved spacetimes, it has been
realised quite early that interactions contribute significantly to particle production due to non-stationary
gravitational forces \cite{BDF}. The scattering
theory of interacting fields on physical backgrounds has thus been in the
focus of attention. The
 picture that has been
drawn is that this scattering behavior is essentially encoded in 
two $S$-matrices, one being a generalisation of the known scattering matrix in flat spacetime that
 can be calculated by means of reduction formulae similar to the Lehmann-Symanzik-Zimmermann (LSZ) method and the other being a
Bogoliubov transformation between "in"- and "out"-representations that are both quasifree
 but not the same \cite{BT}. The former of these $S$-matrices was assumed to describe the scattering of incoming particles into outgoing particles via an interaction with both gravity and matter fields, while the latter, and already well-known one \cite{Wa0}, was interpreted as describing the particle production by means of the gravitational field alone. 

With some exceptions, {\it e.g.}, \cite{BFo}, that, however, have not been
very precise on the topic of asymptotic conditions either, most of
the works within the above described framework \cite{Bi,BD,BDF,BT} have thus assumed asymptotic conditions for the interacting
field in the  Heisenberg picture $\phi(x)\to\phi^\text{in/out}(x)$ as $x^0\to\mp \infty$ with
  $\phi^\text{in}(x)=a(x)+a^\dagger(x)$ and $\phi^\text{out}(x)=b(x)+b^\dagger(x)$, where 
$a(x),\,a^\dagger(x)$ and $b(x),\,b^\dagger(x)$ are annihilation and creation fields fulfilling
$a(x)\Omega^\text{in}=b(x)\Omega^\text{out}=0$ for
suitable "in"- and "out"-Fock, {\it i.e.}, quasifree and pure, 
 vacuum states $\Omega^\text{in}$ and $\Omega^\text{out}$ respectively.  It seems, however, that the consistency of this type of asymptotic condition has
  never been checked.
In fact, once the "in"-state for the field has been specified to be a certain, say,
quasifree state, the field theory should be determined by this initial condition, the canonical commutation relations (CCR),  
 and the equations of motion. Properties of "out"-fields should thus have to be derived
 and cannot be assumed {\it ab initio}.
In \cite{GT}, this supposition has been investigated within a toy
model for quantum field theory on curved spacetimes which has indeed failed to reproduce the assumption of quasifreeness
 of the outgoing state and has thus underlined the relevance of non-quasifree states for free fields.

In the present work, we show that this result is not an artefact of the
mentioned toy model, but occurs genuinely in (scalar) quantum field theory on
non-stationary spacetimes. In fact, we are able to show that the cosmological time scale under which these finding result in non-neglegible particle production is roughly one millonth of a second - early enough not to conflict with well established physical theories like nucleosynthesis, but late enough to be relevant for physical models of the very early universe. 

 This raises the question how to physically 
interpret non-quasifree representations of the CCR, {\it i.e.}, how to compute the particle content of a non-quasifree state. In the case of
unitary equivalence to a quasifree representation and for real scalar quantum fields, we provide a complete solution to this problem using a $\star$-product and the associated calculus on the dual of the Borchers-Uhlmann algebra endowed with a combinatorial co-product. On a Lorentz manifold that is asymptotically flat at early and late times, one can then take
 the distinguished quasifree "in"- and "out"-representations as a reference and calculate the "out"-particle content of the scattered non-quasifree state to finally obtain information on the relative particle production due to interaction with both matter fields and gravitation.
  
In order to be able to access the question whether the outgoing field is in a quasifree representation, one has to explicitely calculate its unsymmetric vacuum expectation values (VEVs), {\it i.e.}, Wightman functions. This is not possible via symmetric scattering amplitudes of LSZ-type or Feynman path integral formalisms, one needs a standalone way to perturbatively compute Wightman functions instead. Such a formalism is
readily available on Minkowski spacetime, namely, the
sectorised Feynman graph
 formalism of Ostendorf and Steinmann \cite{Os,Ste1,Ste2}. This formalism successfully abandons the CCR and asymptotic conditions, but requires the standard spectral properties. It is thus not well-suited for our purposes since it can only be extended to stationary spacetimes, where spectral properties can be formulated, and can not be employed to compute Wightman functions of outgoing fields, but only those of interacting fields.

After fixing the setting of our work in section \ref{setting}, we therefore develop an independent formalism to calculate
the Wightman functions for quantum fields on generic curved spacetimes using a quasifree
 "in"-representation and the perturbation theory of Yang-Feldman equations,
 {\it cf.}, section \ref{calculation_of_wf}. In the course of this, we introduce a Feynman graph calculus for Wightman
 functions where the external vertices of
the Feynman graph have to be labeled properly and the edges have to be
 decorated with arrows symbolising various kinds of propagators.
  Once the method to compute the Wightman functions has been established, their basic properties are then analysed in the
  sections \ref{properties_of_wf} and \ref{locality}. Hermiticity, invariance under
   orthochronous isomorphisms, perturbative positivity, the relation to
asymptotic conditions $\phi(x)\to\phi^\text{out/in}(x)$, $x^0\to\pm\infty$ in the Heisenberg picture, and the spectrum condition (on
stationary spacetimes) do not pose major problems. However, a 
proof of locality by means of our Feynman graphs seems to require a lot of combinatorial effort. We thus take a slight detour: after assuring equivalence of the Yang-Feldman equations and the expansion of the interacting field by means of retarded products, locality follows easily from the Glaser--Lehmann--Zimmermann (GLZ)-relations for retarded products. In this context, the power of the GLZ-relation is sourced from the fact that they seem to be the most efficient way to encode the commutator combinatorics appearing in the proof of locality.

Once we have shown that the Wightman functions in our framework possess all sensible properties we can expect, we proceed in section \ref{prop_out} by verifying the Klein-Gordon equation and the CCR
for the "out"-field. The proof of the CCR is again best achieved by having recourse to retarded products and it is moreover established that the "out"-field representation
is genuinely non-quasifree. To assess the physical content of such a non-quasifree state, we develop the aforementioned method to explicitely calculate a unitary transformation between a non-quasifree state and a quasifree one, provided it exists, in section \ref{unitrafo}. In section \ref{conc}, we provide a brief outlook. Technical and computational details on $\star$-calculus and retarded products are given in two appendices.

We point out that all calculations are carried through in unrenormalised
perturbation theory. This is mostly due to the fact that we have
little to contribute to the theory of renormalisation on physical backgrounds, which has recently been established
by means of the Epstein-Glaser method and microlocal analysis \cite{BFK,BFr,HW1,HW2,HW3}.

\section{Setting}
\label{setting} In this section, we present some basic notation and the field theory setting we work in. To wit, we would like to perturbatively calculate Wightman
functions of hermitian
scalar quantum fields on a globally hyberbolic smooth Lorentzian manifold ($M$,
$g$) in $\phi^p$-theory. That is, our quantum fields, operator valued distributions on a Hilbert space, satisfy the formal\footnote{Recall that we work in unrenormalised formal perturbation theory throughout this paper.} equation
$$\kgo\phi = -\lambda\phi^{p-1}$$
with coupling constant $\lambda$, scalar curvature $R$, mass $m$
and $\mbox{\large{$\sqcup\!\!\!\!\sqcap$}} = \nabla_a \nabla^a$,
$\nabla^a$ being the covariant derivative associated with $g$.

\indent We start by introducing the fundamental functions of the
theory. Let $G_r$ ($G_a$) be the unique \cite{Baer} retarded
(advanced) fundamental solution of the Klein-Gordon operator
$\kgo$, {\it i.e.}, $G_{r/a}$ are real valued bidistributions on $M$
satisfying
$$\kgo G_{r/a}(x,y)=\delta(x,y)$$
and supp$_x$ $G_{r/a}(x,y)\subset \overline V_{y}^{\;\pm}$,
$\overline V_{x}^{\;+}$ ($\overline V_{x}^{\;-}$) being the closed
causal forward (backward) cone with base-point $x$. $\delta(x,y)$ is the
delta-distribution associated with $g$, {\it i.e.}, 
$\int_{M\times M}d_gxd_gy\delta(x,y)f(x)g(y)=\int_Md_gxf(x)g(x)$ for all compactly supported (complex-valued) test functions $f,g\in{\cal D}(M)\dfff C^\infty_0(M,{\mathbb C})$, where
$d_gx=\sqrt{-\bf g}\,dx$, $ {\bf g}=\det g$, is the canonical
volume form associated with $g$.  We note that $G_{r}(x,y) =
G_{a}(y,x)$ and define the antisymmetric bidistribution
\begin{equation}
D(x,y) = G_{r}(x,y) - G_{a}(x,y).\label{def_d}
\end{equation}
Obviously, $D$ fulfils the Klein-Gordon equation in both arguments
and vanishes for $x \perp y$, {\it i.e.}, for spacelike separated $x$
and $y$.

To calculate the Wightman functions, we need to specify
initial conditions for the field $\phi(x)$.
We achieve this by postulating that for large asymptotic times $x^0 \to \mp \infty$
the interacting field $\phi(x)$ converges to incoming or outgoing fields $\phi^{\mbox{\scriptsize{in/out}}}(x)$, where we demand that the "in"-field satisfies the free Klein-Gordon equation $\kgo\phi^{\mbox{\scriptsize{in}}}=0$.
For
space-times $(M,g)$ that are "large" enough to allow the fields to disperse
quickly enough to become finally non-interacting, the abovementioned asymptotic conditions are formulated
in terms of the Yang-Feldman equations \cite{YF}
\begin{equation}
\phi^{\text{loc}}(x)\dfff \phi(x)=
\phi^{\mbox{\scriptsize{in/out}}}(x) + (G_{r/a}\,j)(x),
\label{yf_eq}
\end{equation}
where $(G_{r/a}\,j)(x)$ stands for (the formal expression) 
$$G_{r/a}(x,j)\dfff\int\limits_M d_gy\,G_{r/a}(x,y) j(y)$$
and the current $j$ equals $-\lambda\phi^{p-1}$ in our case.

It is necessary to specify a representation for $\phi^\text{in}(x)$ (or, equivalently, an algebraic "in"-state for the Borchers-Uhlmann algebra of the free scalar field) "by hand" as, in absence of isometries and spectral conditions on general
curved manifolds, the Klein-Gordon equation
 and the CCR are not sufficient to fix it uniquely\footnote{The standard "Fourier" spectrum condition has been successfully replaced by a microlocal spectrum condition \cite{BFK} to advance quantum field theory on curved spacetime in many ways. The microlocal spectrum condition does, however, not determine a unique state, but only a class of states \cite{Ver}.} \cite{Wa}.

This can be accomplished by first choosing propagators $D^{\pm}(x,y)$, that is, complex valued bidistributions on $M$ satisfying the Klein-Gordon
equation in both arguments and Im$D^{\pm}$ = $\pm\frac{1}{2} D$,
$D^{+}(x,y) = \overline{D^{+}(y,x)} = D^{-}(y,x)$, such that $\tilde
D\dfff 2\text{Re}D^{+}$ is symmetric and constitutes the choice of a state. Here, the bar denotes complex comjugation. To select a pure state, we
require $\tilde D(f,f)=\frac{1}{4}\inf_{h\in{\cal
D}(M)}|D(f,h)^2/\tilde D(h,h)|$ for all $f\in{\cal D}(M)$, {\it cf.}, 
\cite{Wa}. Particularly, this implies that $D^{+}$ is positive,
{\it i.e.}, $D^+(\overline f,f) \ge
0$ $\forall$ $f\in{\cal D}(M)$. We
furthermore demand that $D^{+}$ is invariant under any existing isometric
diffeomorphisms of $(M,g)$ preserving the time direction, which only
constrains $\tilde D$, as $D$ automatically fulfils this
condition.  For a discussion of the existence of such (and even
more general) bi-distributions, {\it cf.}, \cite{Wa}.

Before proceeding to select an incoming state, we need to introduce the notion of truncated Wightman functions. These are defined via a cluster
expansion as
\begin{equation}
\langle\Omega,\phi^{a_1}(x_1)\cdots\phi^{a_n}(x_n)\Omega\rangle =
\sum_{I\in\mathcal{P}^{(n)}}\prod_{\{j_1,\cdots,j_l\}\in I}
\langle\Omega,\phi^{a_{j_1}}(x_{j_1})\cdots\phi^{a_{j_l}}(x_{j_l})\Omega\rangle^T,
\label{def_truncated_wf}
\end{equation}
where $a_j\in\{\text{in, loc, out}\}$ and $\mathcal{P}^{(n)}$ is the collection of all partitions of
$\{ 1,\cdots,n\}$ into disjoint, non-empty subsets $\{
j_1,\cdots,j_l\}$ with $j_1 < \cdots < j_l$. A quasifree state is characterised by the property to have vanishing truncated Wightman functions for all $n\neq 2$. 

Let us anticipate at this point that, via the Yang-Feldman equations, both $\phi^\text{loc}$ and $\phi^\text{out}$ are formal power series in $-\lambda$ with monomials in $\phi^\text{in}$ as coefficients and can thus formally be understood as operators on the same, {\it i.e.}, the incoming, Hilbert space, of which $\Omega$ is a vacuum state as we will specify in the following. This motivates our choice to write all Wightman functions as VEVs w.r.t. $\Omega$.

We can now finally specify the state for the incoming field as a quasifree state with two-point function $D^+$, {\it i.e.},  \begin{eqnarray}
\langle\Omega,\phi^{\text{in}}(x)\phi^{\text{in}}(y)\Omega\rangle^T
= D^+(x,y),\;\;\;\;\;\; \label{def_truncated_two_point_in} \\
\label{def_tuncated_n_point_in} \langle\Omega,\phi^{{\rm
in}}(x_1)\cdots\phi^{\text{in}}(x_n)\Omega\rangle^T = 0 \;\;\;
\mbox{for}\; n\not=2. \label{def_truncated_n_point_in}
\end{eqnarray} 

Taking
(\ref{def_truncated_two_point_in})-(\ref{def_truncated_n_point_in})
into account, it follows immediately that (\ref{def_truncated_wf}) simplifies considerably if we only consider Wightman functions of
"in"-fields, namely, 
\begin{equation}
\langle\Omega,\phi^{\text{in}}(x_1)\cdots\phi^{{\rm
in}}(x_n)\Omega\rangle = \left\{
\begin{array}{l l}
 \displaystyle \sum_{I\in\mathcal{P}^{\prime(n)}}\prod_{\{j_1,j_2\}\in I}
D^+(x_{j_1},x_{j_2}) & \quad\mbox{if $n$ is even,}\\\\
 \displaystyle \quad 0 & \quad \mbox{if $n$ is odd.}\\ \end{array}
 \right.\label{cluster_expansion_in}
\end{equation}
Here, $\mathcal{P}^{\prime(n)}$ is the collection of all partitions of
$\{ 1,\cdots,n\}$ into disjoint subsets containing two elements
$\{ j_1,j_2\}$ with $j_1 < j_2$, {\it i.e.}, $\mathcal{P}^{\prime(n)}$ is
a collection of all possible pairings made out of $\{
1,\cdots,n\}$.

We now explain how (\ref{def_truncated_two_point_in}) and (\ref{def_truncated_n_point_in}) determine a particle picture in the remote past. By our assumptions, $\tilde D$ is a symmetric bidistribution that fulfils the Klein-Gordon equation in both
arguments. Consequently, the solution part ${\cal S}f$ of any test function defined as ${\cal S}f\dfff\tilde D\,f$ solves the Klein-Gordon equation and clearly $({\cal S}f,{\cal S}h)\dfff \tilde D(\overline f,g)$
constitutes a well-defined inner product on the space of complex
solutions of the Klein-Gordon equation with compactly supported
initial data. Let us indicate the completion of this space w.r.t. $(\,.\,,\,.\,)$ by ${\cal H}$
and note that the imaginary part $\frac{1}{2}D$ of
$D^+$ defines a ($\C$-bilinear) symplectic form  $\Sigma$ on the
space of complex valued solutions via $\Sigma(Df,Dg)\dfff D(f,g)$
that extends continuously to ${\cal H}$.

Upon comparison with the symplectic form, the inner product then 
induces a complex structure $J$ via $(\psi,J\chi)\dfff\Sigma(\overline
\psi,\chi)$ for any two solutions. One straightforwardly obtains
$J^*=-J$, $J^2=-1$, and, thus, $J=i(K^+-K^-)$ with $K^\pm$ the
projector on the eigenspace of $J$ with eigenvalue $\pm i$. In the following, we call ${\cal H}^\pm\dfff K^\pm{\cal H}$ the
positive/negative frequency spaces respectively. We note that
$\overline{{\cal H}^\pm}={\cal H}^\mp$ since $J\overline\psi=-i\overline \psi$ for $\psi\in{\cal
H}^+$.

Let now ${\cal F}$ be the symmetric Fock space over ${\cal H}^+$ with Fock-vacuum $\Omega$. By $a^\dagger(\psi)$, $a(\psi)$, $\psi\in{\cal H}^+$, we
denote the usual creation and annihilation operators on $\cal F$.
We use the convention $a(\overline\psi)^*=a^\dagger(\psi)^*$,
$\psi\in{\cal H}^+$, in order to obtain a $\C$-linear definition
for $a(\chi)$, $\chi \in{\cal H}^-$. Here, $^*$ stands for taking
the adjoint (neglecting domain questions). Let ${\cal S}_\pm\dfff K^\pm{\cal S}$ be the
operator that maps test functions to the positive/negative frequency solution part.
 The incoming field can now finally be defined as the $\C$-linear
operator valued distribution
\begin{equation}
\label{infield.eqa}
\phi^\text{in}(f)=a({\cal S}_-f)+a^\dagger({\cal S}_+f)\, .
\end{equation}
Furthermore, by \cite[Lemma 3.2.1]{Wa}, $(\tilde Df,\tilde
Dh)=\tilde D(\bar f,h)=\Sigma(\overline{Df},\tilde Dh)=(Df,J\tilde Dh)$ for all
test functions $f,h$, from which $-JDf=\tilde Df$ follows.
Application to (\ref{infield.eqa}) gives $\phi^{\rm
in}(f)=ia(K^-Df)-ia^\dagger(K^+Df)$ which is the definition given in
\cite{Wa}.

It follows from the Fock construction,
(\ref{def_truncated_two_point_in}), and the properties of $D^+$ that
\begin{equation}
 [\phi^{\text{in}}(x), \phi^{\text{in}}(y)] = iD(x,y)
\label{commutator_in}
\end{equation}
which constitutes that the incoming field fulfils the CCR and closes the specification and analysis of the properties of the "in"-field. 

Fixing both $\phi^{\text{in}}(x)$ and $\phi^{\text{out}}(x)$ would
over-determine the system, we therefore only employ the Yang-Feldman equations (\ref{yf_eq}) as
a definition of $\phi^{\text{out}}(x)$ without specifying any further properties of it. From (\ref{def_d}) and (\ref{yf_eq}) it follows immediately that
\begin{equation}
\phi^{\text{out}}(x) = \phi^{\text{in}}(x) + (D\,j) (x).
\label{yf_out}
\end{equation}
Furthermore, because both $\phi^{\text{in}}(x)$ and $D$ fulfil the
Klein-Gordon equation, 
$\phi^{\text{out}}(x)$ does as well. However, we still need to
check if the outgoing field fulfils the CCR and determine whether it is in a quasifree state or not.

\section{Calculation of the Wightman functions}
\label{calculation_of_wf}

To evaluate Wightman functions, we will make use of generalised
Feynman graphs. In the following figures, we draw all graphs
in $\phi^3$-theory for simplicity.
For the actual calculations, the degree $p$ of the
$\phi^p$-theory is irrelevant. We begin developing
the graphical calculus by introducing the symbols for the
propagators of our theory.

\begin{figure}  [htb]\center \includegraphics[width=150pt]{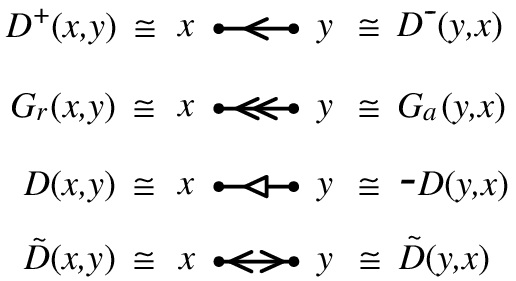} \caption{Propagators} \label{fig_propagators}
\end{figure}

$D^+(x,y)$ is being represented by a line with an open arrow, $D(x,y)$ by a line with a closed arrow and $G_r(x,y)$ by a line
with a double open arrow, the arrows pointing to $x$. Furthermore, $\tilde D(x,y)$ is drawn with two arrows pointing apart as shown in figure \ref{fig_propagators}.

Next, we introduce a tree expansion for the fields according to the
Parisi-Wu method \cite{PW}. We expand the fields in powers of the negative coupling
constant $-\lambda$:

\begin{equation}
\phi^{a}(x) = \sum_{\sigma=0}^{\infty}(-\lambda)^{\sigma}
\phi^{a}_{\sigma}(x).\label{field_expansion}\end{equation}

\noindent Clearly,
\begin{equation}
\phi^{\text{in}}_{\sigma}(x) = \left\{
\begin{array}{l l}
 \phi^{\text{in}}(x) & \quad\mbox{if}\; \sigma=0,\\
 \displaystyle  0 & \quad \mbox{otherwise.}\\ \end{array}\right. \label{field_expansion_in}
\end{equation}
We calculate $\phi^{a}_{\sigma}(x)$ for $a=$ loc/out recursively
using (\ref{yf_eq}) and (\ref{yf_out}):

\begin{equation}
\phi^{a}_{\sigma}(x) = \left\{
\begin{array}{l l}
 \phi^{\text{in}}(x) & \quad\mbox{if}\; \sigma=0 \mbox{ and}\\\\
 \displaystyle \left(\Delta^a \, \sum^{\infty}_{\sigma_1,\cdots,\:\sigma_{p-1}=0,\atop  \sigma_1+\cdots+\sigma_{p-1} =
\sigma-1}\prod_{i=1}^{p-1}\phi^{\text{loc}}_{\sigma_i}\right)(x) & \quad \mbox{otherwise,}\\
\end{array}\right. \label{field_expansion_loc_out}
\end{equation}

\noindent where $\Delta^{\text{loc}}\dfff G_r$, $\Delta^{\text{out}}\dfff D$.
Following the recursion in (\ref{field_expansion_loc_out}), we
define tree graphs corresponding to the summands in
(\ref{field_expansion_loc_out}) by an induction over $\sigma$. To
fix the initial step, we draw $\phi^{a}_{0}(x)$, 
{\it i.e.}, an
"in"-field, as a leaf attached to a root corresponding to an
external $x$-vertex. A tree corresponding to
$\phi^{a}_{\sigma}(x)$ is drawn by taking $p-1$ trees
corresponding to perturbative  local fields of order $\sigma_1,
\cdots, \sigma_{p-1}$ s.t. $\sum \sigma_l = \sigma - 1$,
assembling their roots $y_1, \cdots, y_{p-1}$ to form a single
internal $y$-vertex and adding a trunk, {\it i.e.}, a new line from $y$
to $x$ corresponding to $\Delta^a(x,y)$. Therefore, a tree
correspoding to $\phi^{a}_{\sigma}(x)$ has a root corresponding to
an external $x$-vertex, a trunk corresponding to $\Delta^a$,
several branches corresponding to $G_r$s, several leaves
corresponding to "in"-fields and $\sigma$ branching points
corresponding to internal vertices with a total number of $p-1$
branches and leaves emerging from them. We note that the causal
flow (the direction of the $G_r$-arrows) always points to the
root. 

\begin{figure} [htb]\center \includegraphics[width=250pt]{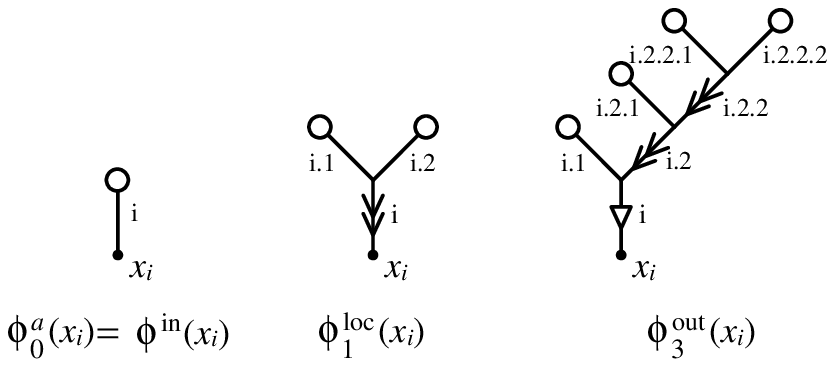} \caption{The only possible tree for $\phi^{a}_{0}(x)$, the only possible tree for $\phi^{\text{loc}}_{1}(x)$, and one possible tree for $\phi^{\text{out}}_{3}(x)$  }\label{fig_trees}  \end{figure}

We label the different tree components inductively,
accounting for the fact that the indices $\sigma_i$ in
(\ref{field_expansion_loc_out}) are distinguishable. The initial
step of the labelling induction is fixed by defining the label of a trunk to
be  the index of the external vertex variable attached to it. To
assign a label to a branch/leaf, one takes the label of the branch
or trunk the considered branch/leaf emerges from as a basis.
One then extends it by a dot followed by a number reflecting the
position of the considered branch/leaf (field) at the actual
branching point (in the corresponding current), {\it i.e.}, the index $i$
of $\phi^{\text{loc}}_{\sigma_i}$ in
(\ref{field_expansion_loc_out}). Some examples of trees are displayed in figure \ref{fig_trees} for the convenience of the reader.

 We shall proceed to consider the Wightman functions of our theory. To compute them, it is
sufficient to consider only their connected parts, {\it i.e.}, the
truncated Wightman functions.

In order to calculate the truncated Wightman functions, we
will first expand them in powers of the coupling constant:
\begin{eqnarray}
\langle\Omega,\phi^{a_1}(x_1)\cdots\phi^{a_n}(x_n)\Omega\rangle^T
=
\sum_{\sigma=0}^{\infty}(-\lambda)^{\sigma}\langle\Omega,\phi^{a_1}(x_1)\cdots\phi^{a_n}(x_n)\Omega\rangle^T_{\sigma}.\label{perturbative_expansion_wf}
\end{eqnarray}
Inserting (\ref{field_expansion}) into the left side of
(\ref{perturbative_expansion_wf}) and comparing terms of equal
order in $-\lambda$, we get:
\begin{eqnarray}
\langle\Omega,\phi^{a_1}(x_1)\cdots\phi^{a_n}(x_n)\Omega\rangle^T_{\sigma}
= \sum^{\infty}_{\sigma_1,\cdots,\:\sigma_n=0,\; \sum \sigma_l =
\sigma}\langle\Omega,\phi^{a_1}_{\sigma_1}(x_1)\cdots\phi^{a_n}_{\sigma_n}(x_n)\Omega\rangle^T.
\label{wf_+_tree_expansion}
\end{eqnarray}
 We know from the tree expansion that, for a fixed
$\sigma$, every $\phi^{a}_{\sigma}(x)$ can be expressed in terms
of incoming fields convoluted with fundamental functions. Therefore, it follows that
$\langle\Omega,\phi^{a_1}_{\sigma_1}(x_1)\cdots\phi^{a_n}_{\sigma_n}(x_n)\Omega\rangle^T$
can be expressed in terms of Wightman functions of "in"-fields integrated with additional propagators.
Finally, combining (\ref{field_expansion_in}),
(\ref{field_expansion_loc_out}), (\ref{cluster_expansion_in}) and
(\ref{wf_+_tree_expansion}), we can express truncated Wightman
functions of arbitrary fields merely in terms of fundamental
functions.

Let us now introduce Feynman graphs corresponding to perturbative
$n$-point Wightman functions. A Feynman graph of order $\sigma$
consists of $n$ external vertices corresponding to the arguments
$x_1, \cdots, x_n$ and type-indices $a_1, \cdots, a_n$ of a
Wightman function and $\sigma$ internal vertices corresponding to
arbitrary variables in $M$. The vertices are connected to the
remainder of the graph by $q$ lines, with $q = 1$ ($q = p$) for
external (internal) vertices. A line is called an external line if
it is connected to an external vertex, an internal line otherwise.
While Wightman functions correspond to Feynman graphs, one can
show that truncated Wightman functions correspond to connected
Feynman graphs. We call a Feynman graph with arrows and labels on
all lines an extended Feynman graph.

On the level of graphs, the resolving of
(\ref{wf_+_tree_expansion}) via (\ref{cluster_expansion_in})
corresponds to gluing the leaves of $n$ trees with a total order
of $\sigma$ together to yield an extended Feynman graph of order
$\sigma$ with $n$ external vertices. As we calculate truncated
Wightman functions, only gluing possibilities that yield connected
Feynman graphs are allowed.

To finish describing the gluing process, we need to analyse the gluing lines in more detail. A line originating from gluing a pair of two leaves together, {\it i.e.},
a $D^{\pm}$-line, is defined to be labelled by combining the
leaves' labels to a pair. Starting from the beginning, we compare
the two labels slot-by-slot until we find a pair of numbers that
does not match. The arrow on the $D^{\pm}$-line then points to the
leaf corresponding to the lower of these numbers.

\begin{figure} [htb]\center \includegraphics[width=350pt]{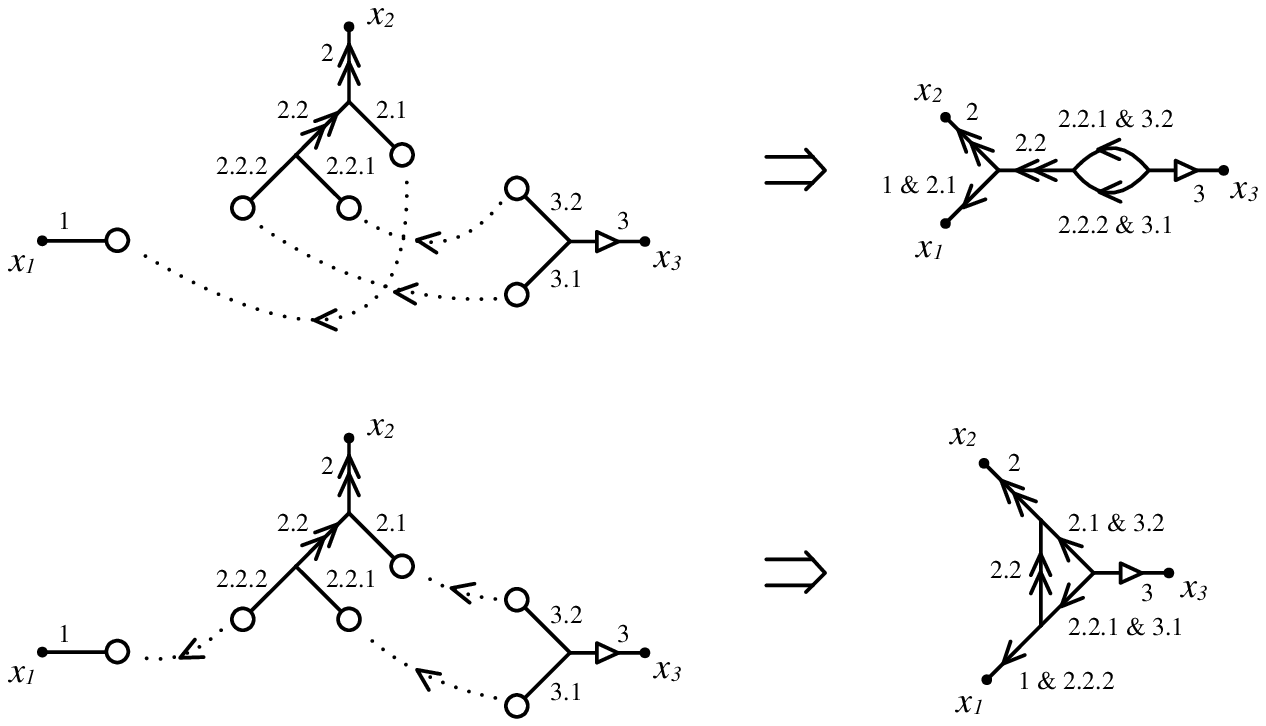} \caption{Two possibilities to glue the same set of trees to graphs
corresponding to $\langle\Omega,\phi^{a}_0(x_1)\phi^{{\rm
loc}}_2(x_2)\phi^{\text{out}}_1(x_3)\Omega\rangle^T$
}\label{fig_gluing_fw}
\end{figure}

\begin{figure} [htb]\center
\includegraphics[width=350pt]{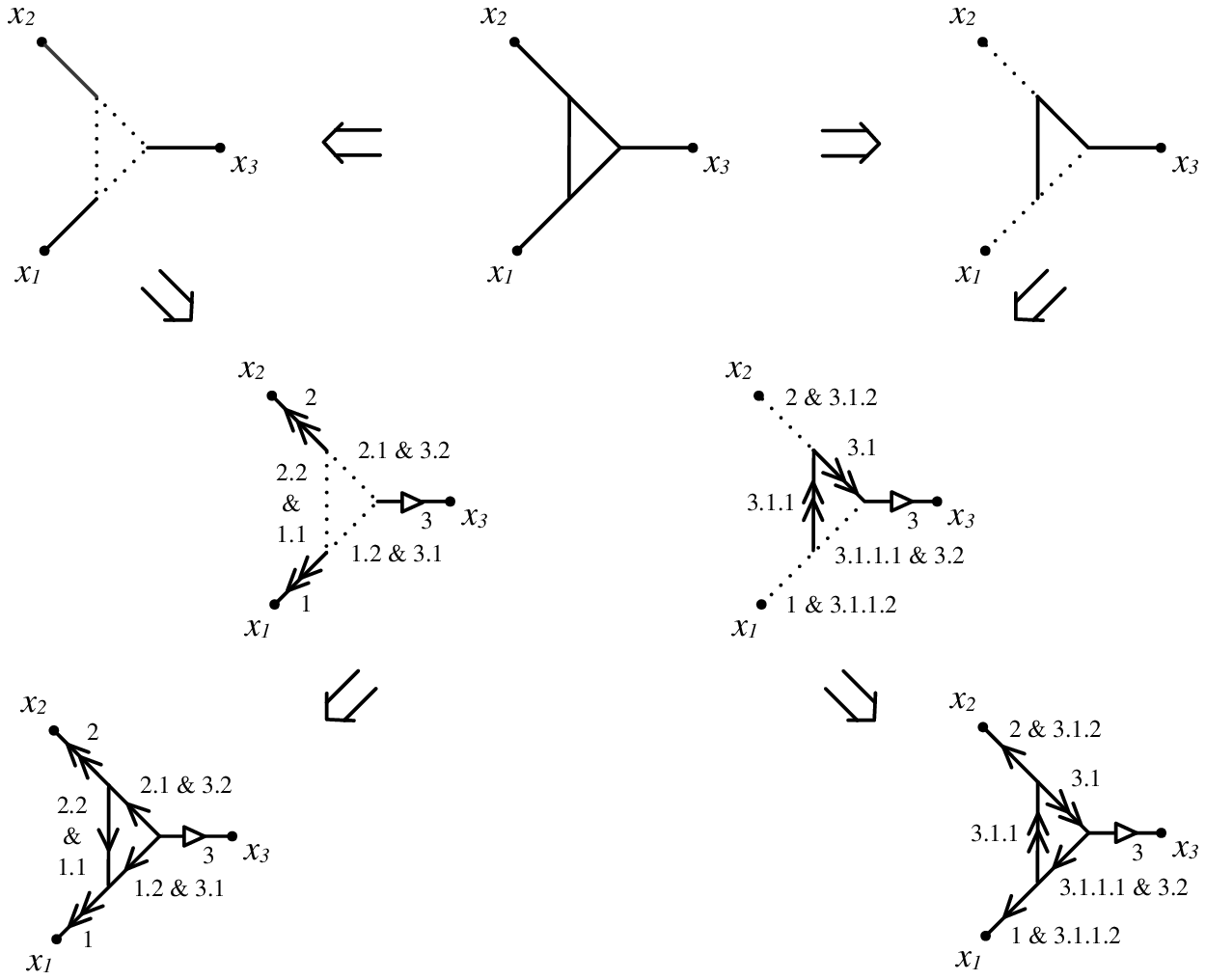} \caption{Two
possibilities to extend a Feynman graph corresponding to
$\langle\Omega,\phi^{\text{loc}}(x_1)\phi^{{\rm
loc}}(x_2)\phi^{\text{out}}(x_3)\Omega\rangle^T_3$
}\label{fig_gluing_bw}
\end{figure}

We have now described how to expand fields to trees that are
subsequently assembled to extended Feynman graphs (see figure
\ref{fig_gluing_fw} for two examples),

\noindent but this also works the other way round. To calculate
$\langle\Omega,\phi^{a_1}(x_1)\cdots\phi^{a_n}(x_n)\Omega\rangle^T_{\sigma}$,
we draw all topologically possible connected Feynman graphs of
order $\sigma$ with $n$ fixed external vertices $x_1, \cdots, x_n$
of type $a_1, \cdots, a_n$. We then consider all possibilities to
partition each Feynman graph into $n$ connected and loop-less
subgraphs, {\it i.e.}, trees and several remaining lines. Each such
subgraph contains exactly one root $x_i$ of type $a_i$. A
partition is fixed by marking a certain number of lines such that the
Feynman graph with these lines removed consists of $n$
disconnected tree graphs without leaves, arrows and labels and
each internal vertex is part of a such a tree (see figure \ref{fig_gluing_bw} for two examples, where the marked lines are displayed as dotted lines). In this process, the external lines
connected to external vertices of type $a=$ in always have to be marked.
Next, we assign labels to all external lines according to the
indices of their external vertex variables. Using these labels as
an initial step and starting from the roots, we "walk up" the trees
on the unmarked lines and inductively assign labels to all lines
emerging from the vertices we pass. As a result, the marked lines
have two labels, one from each vertex they connect. The inductive
dependence of the labels on the preceding labels is fixed by the
labelling algorithm defined above in the discussion of the tree expansion. For each
of the possible choices of labels we draw arrows on all lines. The
type of the arrows on the unmarked lines is chosen according
to $a_1, \cdots, a_n$ and the tree expansion. Furthermore,
 each marked line becomes a $D^{\pm}$-line.
The direction of the arrow on such a line is determined by
comparison of the two labels of the line in the manner described
in the preceding paragraph.

To obtain the analytical expression corresponding to an extended
Feynman graph, we assign variables to all internal vertices, write
down the propagators corresponding to all lines and then integrate
over all internal vertices. Once we have the analytical
expressions, summing over all topologically possible Feynman
graphs and all possibilities to extend them yields
$\langle\Omega,\phi^{a_1}(x_1)\cdots\phi^{a_n}(x_n)\Omega\rangle^T_{\sigma}$
(see figure \ref{fig_gluing_bw} for two examples).

For calculations, it is often convenient to drop
the labels and replace them by combinatorial factors, see \cite{Hack} for a detailed discussion of these issues.

\section{Properties of the Wightman
functions: Invariance, Hermiticity, spectral property, positivity, and the asymptotic condition}

\label{properties_of_wf}

With the means of computing the Wightman functions of our theory at hand, we can continue by discussing their fundamental properties in this section.

\paragraph*{Invariance under orthochronous isometric
diffeomorphisms} As we have shown in section
\ref{calculation_of_wf}, the Wightman functions of our
theory can be expressed in terms of integrals of products
of fundamental functions. Since we know from section \ref{setting}
that all fundamental functions are invariant under isometric
diffeomorphisms preserving the time direction and the integrals
contain the canonical volume form which is invariant under all
isometric diffeomorphisms by definition, invariance of the
Wightman functions follows immediately.

\paragraph*{Hermiticity} The Wightman functions fulfil Hermiticity if
\begin{equation}
\overline{\langle\Omega,\phi^{a_1}(x_1)\cdots\phi^{a_n}(x_n)\Omega\rangle}
=\langle\Omega,\phi^{a_n}(x_n)\cdots\phi^{a_1}(x_1)\Omega\rangle.\label{wf_commu}
\end{equation}
Since we can express the Wightman functions in terms of
Wightman functions of "in"-fields convoluted with the real valued
fundamental functions $G_{r/a}$ and $D$ and since the order of
fields in the latter Wightman function corresponds to the order of
fields in the original Wightman function, it is sufficient to
prove (\ref{wf_commu}) for $a_1,\cdots,a_n=$ in.

We know from (\ref{cluster_expansion_in}) how to express Wightman
functions of incoming fields in terms of $D^+$. The complex
conjugation of (\ref{cluster_expansion_in}) exchanges all
$D^+(x_{j_1},x_{j_2})$ with $D^+(x_{j_2},x_{j_1})$ which obviously
corresponds to reversing the total order of fields in the Wightman
function of "in"-fields, thus (\ref{wf_commu}) holds for
$a_1,\cdots,a_n=$ in.

\paragraph*{Spectral condition} In general
spacetimes, there is no well-defined Fourier transformation,
therefore, standard spectral conditions can not be formulated. However, in
stationary spacetimes, the time translations form a one parameter group
of isometries and we assume that Fourier transformations w.r.t.
the time parameter are possible. A well-defined standard spectral condition
can thus be formulated in that case: one
restricts $\tilde D$ by requiring that the Fourier transform in the time arguments defined by the global
timelike killing field of $n$-point Wightman functions of "in"-fields
vanish if the sums $\Sigma_j\dfff\sum_{j=l+1}^{n}E_l$ are not all positive. Here, 
$E_l$ is the variable conjugated to the $l$-th time argument in
the VEV. We note that all fundamental functions are invariant
under time translations by our assumptions, hence, they only depend on
time differences of both arguments. Therefore, the unitary time
translation operator that is obtained from the time translation
invariance of Wightman functions of interacting and outgoing fields via the GNS construction
coincides with the time translation operator for the "in"-fields,
which has positive spectrum by construction.

\paragraph*{Perturbative positivity} If we expand Wightman functions
perturbatively up to a given order $N$, we can add further terms
of order ${\cal O}(\lambda^{N+1})$ to obtain a VEV of fields
$\phi^{a,N}(x)=\sum_{\sigma=0}^N(-\lambda)^\sigma\phi^{a}_\sigma(x)$
that act as operator-valued distributions on the incoming Fock space.
The VEVs obviously fulfil positivity. Thus, the Wightman
functions expanded in $-\lambda$ up to an arbitrary but fixed order $N$ fulfil positivity
up to a ${\cal O}(\lambda^{N+1})$-term.

\begin{figure} [htb]\center
\includegraphics[width=300pt]{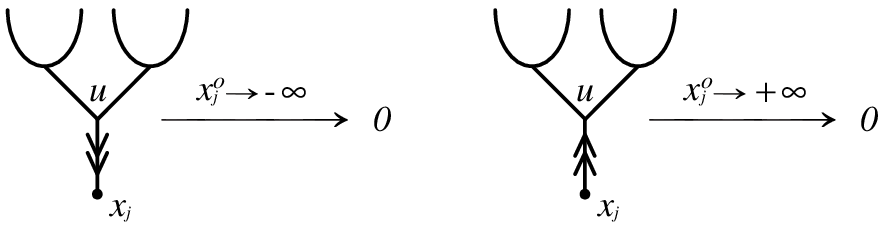} \caption{Heuristic check
of the asymptotic conditions on the level of
graphs}\label{AsympFig}
\end{figure}

\paragraph*{Asymptotic condition} In this work, we have used asymptotic conditions
given by the Yang-Feldman equations (\ref{yf_eq}). It is a
natural question to ask up to what extent these asymptotic conditions lead
to the asymptotic conditions in the Heisenberg picture
$\phi(x)\to\phi^\text{in/out}(x)$ as $x^0\to\mp\infty$ in a given
foliation $M\simeq\R\times {\cal C}$ of $M$, where ${\cal C}$ is a
Cauchy surface. In a weak sense, the Heisenberg asymptotic
condition is
\begin{equation}
\label{Heisi}
\lim_{x_j^0\to\pm\infty}\left[\langle\Omega,\phi^{a_1}(x_1)\cdots\phi(x_j)\cdots\phi^{a_n}(x_n)\Omega\rangle^T-\langle\Omega,\phi^{a_1}(x_1)\cdots\phi^\text{out/in}(x_j)\cdots\phi^{a_n}(x_n)\Omega\rangle^T\right]=0.
\end{equation}
Let us consider the case $x^0\to-\infty$ first: In the extended
Feynman graph expansion of the left hand side, all connected
graphs where the tree with root $x_j$ is of order $\sigma_j=0$ cancel and
all other graphs survive. Likewise, if $x^0\to +\infty$, we obtain
in the expansion into connected extended Feynman graphs two times
all graphs where the order of the $x_j$ tree is larger than zero --
once with the trunk of the $j$-th tree being a retarded propagator
for the local field and once another graph with opposite sign where
the trunk is evaluated with $D$. Using $D=G_{r}-G_{a}$, we see
that, in the case $x^0\to +\infty$, one obtains all extended graphs where the tree with root
$x_j$ is of order larger than zero and its trunk is evaluated with
an advanced propagator, {\it cf.}, figure \ref{AsympFig}. We note that in the
limit $x^0\to\pm\infty$, the integral over the vertex variable $u$
is restricted to $u$ in the causal future/past of $x_j$ and hence
the domain of integration becomes smaller and smaller. An actual
proof of the vanishing of the left hand side of (\ref{Heisi})
requires technical assumptions on the manifold $M$ and on the
propagators $D^\pm$ -- and hence the state -- and we do not want
to go into the details now. It however seems that these conditions
are not much stronger than what is needed to assure that the
integrals over the vertices in the Feynman graphs exist.

\section{Properties of the Wightman
functions: Locality } \label{locality}

To prove locality of the truncated Wightman functions, we have to
show that
\begin{equation}
\langle\Omega,\phi^{a_1}(x_1)\; ...\;
[\phi^{a_i}(x_i),\phi^{a_{i+1}}(x_{i+1})]\cdots
\phi^{a_n}(x_n)\Omega\rangle^T \label{wf_locality}
\end{equation}
 vanishes for $x_i \perp x_{i+1}$,$\;\forall i\in\{1,\;...\;,n-1\},\;a_j=$ loc if
$j\in\{i,i+1\},\;a_j=$ in/loc/out otherwise. For proving this, it is sufficient to show that the interacting field itself is local. Since we are given the
interacting field as a formal power series in $-\lambda$, locality of $\phi=\phi^\text{loc}$ has to be proven to each order in $-\lambda$ separately. To order
$\sigma$ we have

\begin{equation}
    \left[\phi(x),\phi(y)\right]_\sigma = \sum\limits_{\sigma_1+\sigma_2=\sigma} \left[\phi(x)_{\sigma_1},\phi(y)_{\sigma_2}\right].
\end{equation}

\begin{figure} [htb]\center
\includegraphics[width=300pt]{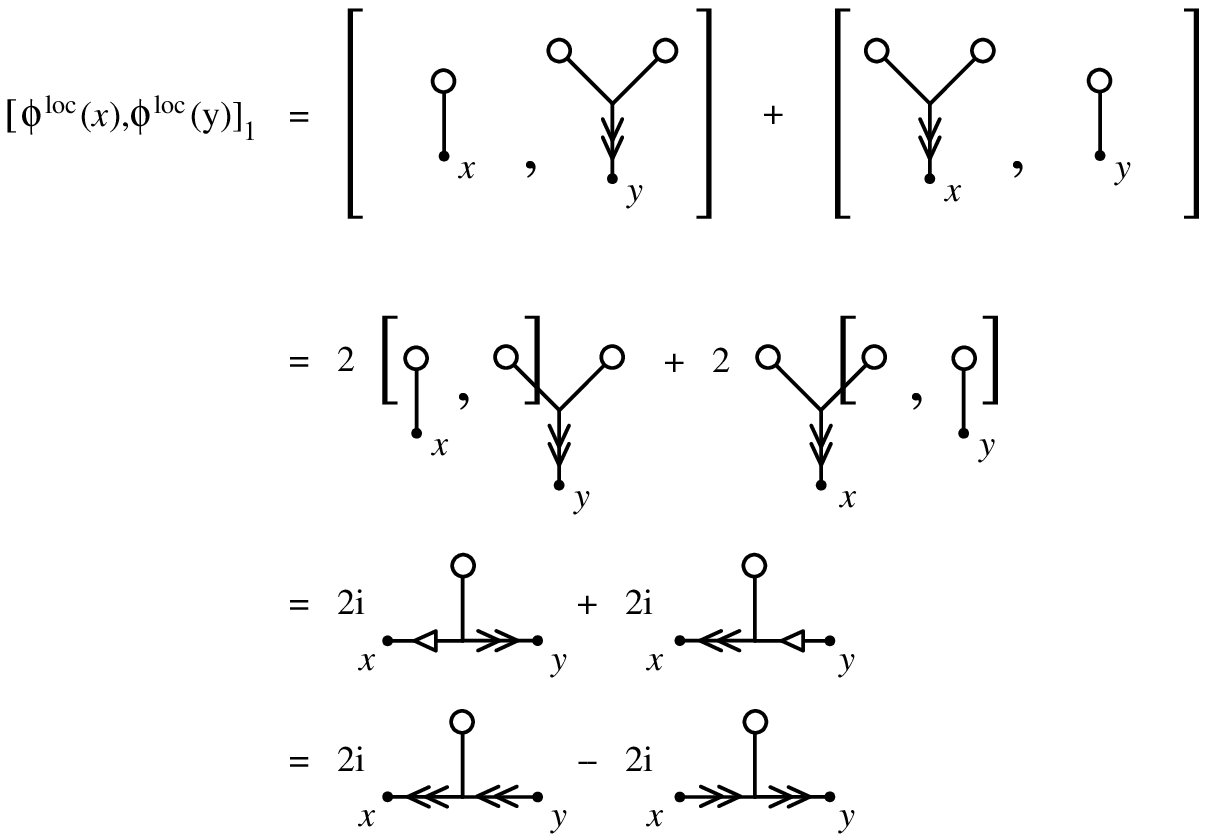}\caption{The commutator of two interacting fields to first order}\label{commu1}
\end{figure}

For a fixed order, it is hence possible to replace the fields $\phi(x)_{\sigma_1}$, $\phi(x)_{\sigma_2}$ by their tree expansions and
compute $\left[\phi(x),\phi(y)\right]_\sigma$ as sums of commutators of single trees that are in effect commutators of products of free
fields integrated with retarded propagators. Employing Leibniz' rule for the commutator results in glueing together two leaves, one from
each tree, with a $D$-propagator. One can then hope to obtain an expression which
vanishes for spacelike-separated $x$ and $y$. The procedure for $\sigma=1$ in $\phi^3$-theory is depicted in figure \ref{commu1}, where the last step follows from
$D=G_r-G_a$ and "telescope cancellations".

It turns out that one is left with a
"$G_r$-chain" and a "$G_a$-chain", {\it i.e.}, a product of retarded/advanced propagators such that the right slot of one propagator
corresponds to the left slot of another, namely,

\begin{equation}
    \left[\phi(x),\phi(y)\right]_1 = 2i\int\limits_M\,d_gx_1\,\left\{G_r(x,x_1)G_r(x_1,y)- G_a(x,x_1)G_a(x_1,y)\right\} \phi^{\text{in}}(x_1).
\end{equation}

Owing to the causal support properties of the retarded and advanced propagators, we see that either $x\preceq x_1 \preceq y$ or $x\succeq
x_1 \succeq y$, where $x \preceq y$ $(x \succeq y)$ depicts that $x\in \overline V^\pm_y$. As a result, both the $G_r$-chain
and the $G_a$-chain vanish for $x \perp y$. To generalise this observation to arbitrary order, we prove the equivalence of the tree expansion to the expansion into retarded products. We then use the GLZ relation \cite{GLZ} to prove locality. See also \cite{Hack} for a proof up to second loop order that works on the level of Feynman graphs and proves locality graph by graph. 

The retarded product $R_{1,n}(B_0(x_0)\,|\, B_1(x_1),\dots,B_n(x_n))$ of $n+1$ operators $B_0(x_0)$, $\dots$, $B_n(x_n)$ that are mutually local, {\it i.e.}, $[B_i(x_i),B_j(x_j)]$ vanishes for all $i$, $j$ if $x_i\perp x_j$, is defined as

$$R_{1,0}\left(B_0(x_0)\right)\dfff B_0(x_0),$$
$$R_{1,n}\left(B_0(x_0)\,|\, B_1(x_1),\dots,B_n(x_n)\right)\dfff$$
\begin{equation}
 \label{defret}
   \dfff (-1)^n\sum\limits_{\pi \in S_n}\left[B_{\pi(n)}(x_{\pi(n)}), \left[B_{\pi(n-1)}(x_{\pi(n-1)}), \cdots \left[B_{\pi(1)}(x_{\pi(1)}),B_0(x_0)\right]\cdots\right]\right]  {\bf 1}_{x_0\preceq x_{\pi(1)} \preceq \cdots \preceq x_{\pi(n)}},
\end{equation}
where $S_n$ is the permutation group of order $n$ and $ {\bf 1}_A$ denotes the characteristic function of the set $A$. Note that renormalisation is necessary for a proper definition at coinciding points. With the above definition, $R_{1,n}(B_0(x_0)\,|\, B_1(x_1),\dots,B_n(x_n))$ is both symmetric in the last $n$ slots and manifestly vanishing if any of the spacetime positions of the last $n$ operators is not in the causal past of $x_0$. From \eqref{defret} it follows straighforwardly that retarded products of a certain order may be expressed in terms of retarded products of one order less 
$$R_{1,n}\left(B_0(x_0)\,|\, B_1(x_1),\dots,B_n(x_n)\right)=$$
\begin{equation}
 \label{retrecursion1}
    = -\sum\limits_{j=1}^n\left[B_j(x_j), R_{1,n-1}(B_0(x_0)\,|\, B_1(x_1),\dots,\cancel{ B_j(x_j)},\dots,B_n(x_n))\right] {\bf 1}_{x_j\succeq x_i\;\forall i\in\{0,\dots,n\}}.
\end{equation}
Employing the above recursion and the Jacobi identity for the commutator, one can prove the GLZ relations \cite{GLZ}, 
$$R_{1,n}\left(A\,|\, C, B_1(x_1),\dots,B_n(x_n)\right)-R_{1,n}\left(C\,|\, A, B_1(x_1),\dots,B_n(x_n)\right)=$$
\begin{equation}
 \label{glz}
   =\sum\limits_{I\subset N\df\{1,\dots,n\}}\left[R_{1,|I|}\left(A\,\left|\,\prod\limits_{i\in I}B_i(x_i)\right.\right),R_{1,|N\setminus I|}\left(C\,\left|\prod\limits_{j\in N\setminus I}B_j(x_j)\right.\right)\right].
\end{equation}

The most important feature of retarded products in our context is the fact that the interacting field can be expanded in terms of retarded products, where a retarded product of order $\sigma$ encodes the complete $(-\lambda)^\sigma$-contribution of the interacting field, namely,
\begin{equation}
 \label{retexpansion}
   \phi(x)=\sum\limits_{\sigma=0}^\infty\frac{(i\lambda)^\sigma}{\sigma!}\int\limits_{M^\sigma}\prod\limits_{i=1}^\sigma d_gx_i \; R_{1,\sigma}(\phi^\text{in}(x)\,|\,{\mathcal L}_\text{int}(x_1),\dots,{\mathcal L}_\text{int}(x_\sigma)),
\end{equation}
with ${\mathcal L}_\text{int}(x)\dfff \phi^\text{in}(x)^p/p$ in our case. A combinatorial proof of (\ref{retexpansion}) that makes the relation with the tree expansion explicit will be given at the end of this section.  

To prove locality in terms of retarded products, we start from 
$$\left[\phi(x),\phi(y)\right]_\sigma = \int\limits_{M^\sigma}\prod\limits_{i=1}^\sigma d_gx_i\sum\limits_{\sigma_1+\sigma_2=\sigma}\frac{(-i)^{\sigma}}{\sigma_1!\sigma_2!}\;\times$$$$\times\;
\left[ R_{1,\sigma_1}(\phi^\text{in}(x)\,|\,{\mathcal L}_\text{int}(x_1),\dots,{\mathcal L}_\text{int}(x_{\sigma_1})),  R_{1,\sigma_2}(\phi^\text{in}(y)\,|\,{\mathcal L}_\text{int}(x_{\sigma_1+1}),\dots,{\mathcal L}_\text{int}(x_{\sigma}))\right]$$
which, due to the GLZ relations, simplifies to
$$\left[\phi(x),\phi(y)\right]_\sigma = \frac{(-i)^{\sigma}}{\sigma!}
\left\{R_{1,\sigma+1}(\phi^\text{in}(x)\,|\,\phi^\text{in}(y), {\mathcal L}_\text{int}^{\otimes\sigma})-  R_{1,\sigma+1}(\phi^\text{in}(y)\,|\,\phi^\text{in}(x), {\mathcal L}_\text{int}^{\otimes\sigma})\right\},$$ where $R_{1,n+1}\left(\phi^\text{in}(x)\,|\,\phi^\text{in}(y), {\mathcal L}_\text{int}^{\otimes n}\right)$ stands in shorthand for $$\int\limits_{M^n}\prod\limits_{i=1}^{\sigma} d_gx_i R_{1,n+1}(\phi^\text{in}(x)\,|\,\phi^\text{in}(y), {\mathcal L}_\text{int}(x_1),\dots,{\mathcal L}_\text{int}(x_n)).$$ The result is, as we could have expected from our finger exercise in figure \ref{commu1}, a retarded piece, presumably corresponding to a sum of $G_r$-chains, minus an advanced piece, presumably corresponding to a sum of $G_a$-chains. In fact, we will prove this correspondence in appendix \ref{retapp}, since it will be necessary for the proof of the "out"-field CCR. If $x\perp y$, both the retarded and the advanced piece vanish due to the causal support properties of the retarded products.

Let us now proceed to see why the perturbative expansion of the interacting field in terms of retarded products is equivalent to the perturbative expansion of the interacting field due to the Yang-Feldman equation in terms of tree graphs. Since in case of the expansion in trees the $(-\lambda)^\sigma$-contribution to the interacting field consists of the sum of all possible tree graphs with $\sigma$ branching points, we have to show that the $(-\lambda)^\sigma$ term of the expansion in retarded products corresponds to exactly such sum. To achieve this, let us recall how one can inductively obtain all possible trees of order $\sigma$: one starts with the "in"-field and then replaces an "in"-field/leaf by a first order-tree $\sigma$ times and in all possible ways, taking care to discard trees occuring multiply. Starting from the recursion \eqref{retrecursion1}, we shall proceed to understand how it reproduces the aforementioned combinatorial procedure.

To this effect, let us first notice that \eqref{retrecursion1} simplifies considerably in the current context, since all operators $B_1,\dots,B_n$ are of the same type and the labelling of the $x_i$ does not matter due to them being integration variables. Hence, the symmmetrisation can be replaced by a factor and the $(-\lambda)^\sigma$-contribution to the interacting field in terms of retarded products reads

\begin{align}\phi(x)_{\sigma}&=\frac{(-i)^{\sigma}}{\sigma!}\int\limits_{M^\sigma}\prod\limits_{i=1}^{\sigma} d_gx_i \; R_{1,\sigma}(\phi^\text{in}(x)\,|\,{\mathcal L}_\text{int}(x_1),\dots,{\mathcal L}_\text{int}(x_\sigma))\notag\\
&=-\frac{(-i)^{\sigma}}{(\sigma-1)!}\int\limits_{M^\sigma}\prod\limits_{i=1}^{\sigma} d_gx_i \left[{\mathcal L}_\text{int}(x_{\sigma}),R_{1,\sigma-1}(\phi^\text{in}(x)\,|\,{\mathcal L}_\text{int}(x_1),\dots,{\mathcal L}_\text{int}(x_{\sigma-1}))\right]{\bf 1}_{x_\sigma\succeq x_i\atop\forall i\in\{0,\dots,\sigma-1\}}\notag.
\end{align}

Since all operators appearing in $R_{1,\sigma-1}(\phi^\text{in}(x)\,|\,{\mathcal L}_\text{int}(x_1),\dots,{\mathcal L}_\text{int}(x_{\sigma-1}))$ are monomials in the "in"-field, we know due to the Leibniz rule for the commutator that this retarded product is given by sums of products of "in"-fields. As ${\mathcal L}_\text{int}(x_{\sigma})$ is also a monomial in the incoming field and the commutator of two "in"-fields is a C-number, we can compute $[{\mathcal L}_\text{int}(x_{\sigma}),$$R_{1,\sigma-1}(\phi^\text{in}(x)\,|\,{\mathcal L}_\text{int}(x_1),\dots,{\mathcal L}_\text{int}(x_{\sigma-1}))]$ by means of
$$\left[B^n,\prod\limits_{i=1}^m A_i\right]=\sum\limits_{j=0}^{m}\prod\limits_{i=1}^{j-1} A_i \;[B,A_j]\frac{dB^{n}}{dB} \prod\limits_{i=j+1}^m A_i,$$ which holds under the assumption that $[A_i,B]$ commutes with all other occuring operators for all $i$  and where we have formally written $dB^{n}/dB$ in shorthand for $nB^{n-1}$. This formula implies that $[{\mathcal L}_\text{int}(x_{\sigma}),$$R_{1,\sigma-1}(\phi^\text{in}(x)\,|\,{\mathcal L}_\text{int}(x_1),\dots,{\mathcal L}_\text{int}(x_{\sigma-1}))] {\bf 1}_{x_\sigma\succeq x_i\forall i\in\{0,\dots,\sigma-1\}}$ can be computed by summing over all possibilities to replace one incoming field $\phi^\text{in}(x_j)$ in $R_{1,\sigma-1}(\phi^\text{in}(x)\,|\,{\mathcal L}_\text{int}(x_1),\dots,{\mathcal L}_\text{int}(x_{\sigma-1}))$ by

\begin{align}
\left[\phi^\text{in}(x_\sigma),\phi^\text{in}(x_j)\right]\frac{d{\mathcal L}_\text{int}(x_{\sigma})}{d\phi^\text{in}(x_\sigma)} {\bf 1}_{x_\sigma\succeq x_j}&=iD(x_\sigma, x_j)\phi^\text{in}(x_\sigma)^{p-1} {\bf 1}_{x_\sigma\succeq x_j}\notag\\&=-iG_r(x_j,x_\sigma)\phi^\text{in}(x_\sigma)^{p-1}.\notag
\end{align}

To account for the Leibniz rule of this procedure, we denote it by means of a formal derivative operator ({\it cf.}, \cite{Duetsch}, where the framework is such that it is a well-defined functional derivative), {\it viz.},
$${\mathcal D}(x)\phi^\text{in}(y)\dfff G_r(\,\cdot\,,x)\frac{d{\mathcal L}_\text{int}(x)}{d\phi^\text{in}(x)}\frac{d}{d\phi^\text{in}(\,\cdot\,)}\phi^\text{in}(y)\dfff G_r(y, x)\frac{d{\mathcal L}_\text{int}(x)}{d\phi^\text{in}(x)}.$$
Employing this notation, we can write the interacting field to order $\sigma$ as

\begin{align}\phi(x)_{\sigma}&=-\frac{(-i)^{\sigma}}{(\sigma-1)!}\int\limits_{M^\sigma}\prod\limits_{i=1}^{\sigma} d_gx_i \; \left[{\mathcal L}_\text{int}(x_{\sigma}),R_{1,\sigma-1}(\phi^\text{in}(x)\,|\,{\mathcal L}_\text{int}(x_1),\dots,{\mathcal L}_\text{int}(x_{\sigma-1}))\right]{\bf 1}_{x_\sigma\succeq x_i\atop\forall i\in\{0,\dots,\sigma-1\}}\notag\\&=\frac{(-i)^{\sigma-1}}{(\sigma-1)!}\int\limits_{M^\sigma}\prod\limits_{i=1}^{\sigma} d_gx_i \; {\mathcal D}(x_{\sigma})R_{1,\sigma-1}(\phi^\text{in}(x)\,|\,{\mathcal L}_\text{int}(x_1),\dots,{\mathcal L}_\text{int}(x_{\sigma-1})){\bf 1}_{x_\sigma\succeq x_i\atop\forall i\in\{0,\dots,\sigma-1\}}\notag,
\end{align} where it is understood that the operator expression replacing the "in"-field on which ${\mathcal D}(x_{\sigma})$ currently acts has to be inserted exactly at the position where that incoming field has been. We can now iterate the recursion \eqref{retrecursion1} to eventually obtain
\begin{equation}\label{pretrees}\phi(x)_{\sigma}=\int\limits_{M^\sigma}\prod\limits_{i=1}^{\sigma} d_gx_i \; {\mathcal D}(x_\sigma)\cdots{\mathcal D}(x_1)\phi^\text{in}(x) {\bf 1}_{x_\sigma\succeq \cdots \succeq x_1 \succeq x}.\end{equation}

This puts us in the position to prove formal equivalence of this expression to the one obtained from the Yang-Feldman equation in terms of tree graphs. Obviously, both expression are equal in zeroth order. Let us assume that they are equal to order $\sigma$ and show how equivalence to order $\sigma+1$ follows inductively. By our assumptions, the integrand of \eqref{pretrees} can be rewritten in a form devoid of explicit restrictions on the integration domain for $x_1, \dots, x_\sigma$, namely, the sum of the integrands of all trees of order $\sigma$, {\it viz.}, $$\phi(x)_{\sigma}\fd\int\limits_{M^\sigma}\prod\limits_{i=1}^{\sigma} d_gx_i \;I_\sigma(x_1, \dots, x_\sigma).$$ With this notation, we have $$\phi(x)_{\sigma+1}=\int\limits_{M^{\sigma+1}}\prod\limits_{i=1}^{\sigma+1} d_gx_i {\mathcal D}(x_{\sigma+1})I_\sigma(x_1, \dots, x_\sigma) {\bf 1}_{x_{\sigma+1}\succeq x_i\atop\forall i\in\{0,\dots,\sigma\}}.$$ ${\mathcal D}(x_{\sigma+1})$ acts on $I_\sigma$ by replacing an "in"-field by the integrand of a first order tree graph in all tree integrands $I_\sigma$ consists of and in all possible ways. As a result, we obtain integrands of trees of order $\sigma+1$, still with no mutual restrictions on the integration variables $x_1, \dots, x_\sigma$, but with the constraint that $x_{\sigma+1}$ is causally earlier than all of these spacetimes points. It is clear that ${\mathcal D}(x_{\sigma+1})I_\sigma(x_1, \dots, x_\sigma){\bf 1}_{x_{\sigma+1}\succeq x_i \forall i\in\{0,\dots,\sigma\}}$ contains all possible tree integrands of order $\sigma+1$, the question is whether on can rewrite this expression in way that it contains every possible tree integrand of order $\sigma+1$ with unit weight and without any restrictions on the integration domain. 

To answer this question, let us divide the tree integrands we straightforwadly obtain by the action of ${\mathcal D}(x_{\sigma+1})$ on $I_\sigma(x_1, \dots, x_\sigma)$ into equivalence classes, where two integrands are taken to be equivalent if they can be matched by permuting vertex variables. Let us choose an arbitrary but fixed equivalence class $E$ and assume that it has $m$ members. Their number implies that there are $m$ different possibilities to build the tree graph corresponding to $E$ out of a tree graph of order $\sigma$ by replacing a leaf with a first order tree. Hence, the tree graph corresponding to $E$ must have $m$ "virgin" vertices, where we call a vertex virgin if it has the maximal number of $p-1$ "in"-fields attached to it; the $m$ members of $E$ then correspond to the $m$ possible ways to remove one of these virgin vertices to obtain a tree of lower order. Now, let us permute the vertex variables of all members of $E$ such that they are all equal but have different restrictions on the integration domain of $x_1, \dots, x_{\sigma+1}$ and let ${\mathcal V}\dfff \{x_{i_1},\dots,x_{i_m}\}$ be the resulting integration variables of the virgin vertices. Since the virgin vertices are connected to the remainder of the tree by retarded propagators, we can discard the integration constrains on the $m$ integrands under consideration which are automatically fulfilled due to the causal support properties of $G_r$. The remaining restrictions on the integration domain can only be of the form ${\mathcal V}\ni x_j\succeq x_i \;\forall x_i\in {\mathcal V}$. In fact, due to the Leibniz rule for ${\mathcal D}(x_{\sigma+1})$, all $m$ possible integration constraints of this kind must appear. Hence, summing up the $m$ integrands with matching variables but different restrictions on the integration domain, we obtain once the same integrand, but without any integration constraints. Since the above described procedure is valid for all equivalence classes of tree integrands, $\int_{M^{\sigma+1}}\prod_{i=1}^{\sigma+1} d_gx_i {\mathcal D}(x_{\sigma+1})\cdots{\mathcal D}(x_1)\phi^\text{in}(x) {\bf 1}_{x_{\sigma+1}\succeq \cdots \succeq x_1 \succeq x}$ corresponds to the sum of all possible trees of order $\sigma+1$ weighted with unit multiplicity, and the formal equivalence of the expansion of the interacting field by means of the retarded products on the one hand and the Yang-Feldman equation via Parisi-Wu tree graphs on the other hand is established.

\section{Properties of the outgoing fields}\label{prop_out}

We now examine the properties of the outgoing fields.

\paragraph*{Klein-Gordon Equation}
We have already seen in section \ref{setting} that the
"out"-fields of the theory fulfil the Klein-Gordon equation as is manifest from their definition via the Yang-Feldman equation \eqref{yf_out}.

\paragraph*{Canonical commutation relations}

To prove the CCR, we have to examine the commutator of two outgoing fields to each order in $-\lambda$ separately. Due to the Yang-Feldman equations, we have
$$\left[\phi^\text{out}(x),\phi^\text{out}(y)\right]_\sigma=\left[\phi(x)-(-\lambda)G_a(x,
\phi^{p-1}),\phi(y)-(-\lambda)G_a(y,
\phi^{p-1})\right]_\sigma$$
\begin{equation}\label{retccr}=\underbrace{\left[\phi(x),\phi(y)\right]_\sigma}_{I_\sigma}-\underbrace{\left[\phi(x),G_a(y,
\phi^{p-1})\right]_{\sigma-1}}_{II_\sigma}-\underbrace{\left[G_a(x,
\phi^{p-1}),\phi(y)\right]_{\sigma-1}}_{III_\sigma}+\underbrace{\left[G_a(x,
\phi^{p-1}),G_a(y,
\phi^{p-1})\right]_{\sigma-2}}_{IV_\sigma}.\end{equation} In zeroth order, only the first term of \eqref{retccr} contributes and the result is $$\left[\phi^\text{out}(x),\phi^\text{out}(y)\right]_0=\left[\phi^\text{in}(x),\phi^\text{in}(y)\right]=iD(x,y).$$ To prove CCR for the "out"-field, we thus need to show that $\left[\phi^\text{out}(x),\phi^\text{out}(y)\right]_\sigma$ vanishes identically for $\sigma > 0$. In the case $\sigma=1$, only the first three terms of \eqref{retccr} contribute and we have

\begin{align}
I_1 &= -i\int\limits_M\,d_gx_1\,\left\{R_{1,2}\left(\phi^\text{in}(x)\,|\,\phi^\text{in}(y), {\mathcal L}_\text{int}(x_1)\right)- R_{1,2}\left(\phi^\text{in}(y)\,|\,\phi^\text{in}(x), {\mathcal L}_\text{int}(x_1)\right)\right\}\notag\\
&=(p-1)i\int\limits_M\,d_gx_1\,\left\{G_r(x,x_1)G_r(x_1,y)- G_a(x,x_1)G_a(x_1,y)\right\} \phi^{\text{in}}(x_1)^{p-2}\notag\\
II_1 &= \left[\phi^\text{in}(x), G_a(y, (\phi^\text{in})^{p-1})\right]= (p-1)i\int\limits_M\,d_gx_1\,D(x,x_1)G_r(x_1,y)\phi^\text{in}(x_1)^{p-2}\notag\\
III_1 &= \left[G_a(x, (\phi^\text{in})^{p-1}), \phi^\text{in}(y)\right]= (p-1)i\int\limits_M\,d_gx_1\,G_a(x,x_1)D(x_1,y)\phi^\text{in}(x_1)^{p-2}\notag.
\end{align}Due to "telescope cancellations" by means of $D=G_r-G_a$, $I_1-II_1-III_1=0$.

For $\sigma>1$, the structure of cancellations is in principle the same as above, the only difference being the possible appearance of propagators "lying inbetween" $G_{r/a}(x,x_1)$ and $G_{r/a}(x_1,y)$. Such terms survive in $I_\sigma-II_\sigma-III_\sigma$ and have to be cancelled by $IV_\sigma$. To treat such terms, we need the following identities, proven in appendix B:

$$\frac{(-i)^n}{n!}R_{1,n+1}\left(\phi^\text{in}(x)\,|\,\phi^\text{in}(y), {\mathcal L}_\text{int}^{\otimes n}\right)$$
\begin{equation}\label{retpull1}=\int\limits_M d_gx_1\;G_r(x,x_1)\sum\limits_{\sum \sigma_i = n-1}\sum\limits_{j=0}^{p-2}\phi(x_1)^j_{\sigma_1}\,\frac{(-i)^{\sigma_2}}{\sigma_2!}R_{1,\sigma_2+1}\left(\phi^\text{in}(x_1)\,|\,\phi^\text{in}(y), {\mathcal L}_\text{int}^{\otimes \sigma_2}\right)\,\phi(x_1)^{p-2-j}_{\sigma_3}\end{equation}
\begin{equation}\label{retpull2}=\int\limits_M d_gx_1\sum\limits_{\sum \sigma_i = n-1}\sum\limits_{j=0}^{p-2}\phi(x_1)^j_{\sigma_1}\,\frac{(-i)^{\sigma_2}}{\sigma_2!}R_{1,\sigma_2+1}\left(\phi^\text{in}(x)\,|\,\phi^\text{in}(x_1), {\mathcal L}_\text{int}^{\otimes \sigma_2}\right)\,\phi(x_1)^{p-2-j}_{\sigma_3}\;G_r(x_1,y),\end{equation}
where $\phi(x)^j_\sigma$ stands in shorthand for $\sum\limits_{\sum\sigma_i=\sigma}\prod\limits_{i=1}^j\phi(x)_{\sigma_i}$.
Iterating these identies yields that $\frac{(-i)^n}{n!}R_{1,n+1}\left(\phi^\text{in}(x)\,|\,\phi^\text{in}(y), {\mathcal L}_\text{int}^{\otimes n}\right)$ can be expressed completely in terms of $G_r$-chains connecting $x$ with $y$.

Employing the above listed identities, we have for $\sigma > 1$

\begin{align}
I_\sigma &= \frac{(-i)^\sigma}{\sigma!}\left\{R_{1,\sigma+1}\left(\phi^\text{in}(x)\,|\,\phi^\text{in}(y), {\mathcal L}_\text{int}^{\otimes \sigma}\right)-R_{1,\sigma+1}\left(\phi^\text{in}(y)\,|\,\phi^\text{in}(x), {\mathcal L}_\text{int}^{\otimes \sigma}\right)\right\}\notag\\
&=\int\limits_{M}d_gx_1 \sum\limits_{\sum\sigma_i=\sigma-1}\sum\limits_{i=0}^{p-2}\phi(x_1)^i_{\sigma_1}\frac{(-i)^{\sigma_2}}{\sigma_2!}
 \left\{G_r(x,x_1)R_{1,\sigma_2+1}\left(\phi^\text{in}(x_1)\,|\,\phi^\text{in}(y), {\mathcal L}_\text{int}^{\otimes \sigma_2}\right)-\right.\notag\\
 &\qquad\left.-\;G_a(x,x_1)R_{1,\sigma_2+1}\left(\phi^\text{in}(y)\,|\,\phi^\text{in}(x_1), {\mathcal L}_\text{int}^{\otimes \sigma_2}\right)\right\}\phi(x_1)^{p-2-i}_{\sigma_3}\notag\\
&=(p-1)i\int\limits_{M}d_gx_1 \left\{G_r(x,x_1)G_r(x_1,y)-G_a(x,x_1)G_a(x_1,y)\right\}\phi(x_1)^{p-2}_{\sigma-1}\;+\notag\\
 &\qquad+\int\limits_{M^2}d_gx_1 d_gx_2\sum\limits_{\sum\sigma_i=\sigma-2}\sum\limits_{i,j=0}^{p-2}\phi(x_1)^i_{\sigma_1}\phi(x_2)^j_{\sigma_2}\frac{(-i)^{\sigma_3}}{\sigma_3!}\times\notag\\&\qquad\qquad\times\left\{G_r(x,x_1)R_{1,\sigma_3+1}\left(\phi^\text{in}(x_1)\,|\,\phi^\text{in}(x_2), {\mathcal L}_\text{int}^{\otimes \sigma_3}\right)G_r(x_2,y)\;-\right.\notag\\&\qquad\qquad
\left.-\;G_a(x,x_1)R_{1,\sigma_3+1}\left(\phi^\text{in}(x_2)\,|\,\phi^\text{in}(x_1), {\mathcal L}_\text{int}^{\otimes \sigma_3}\right)G_a(x_2,y)\right\}\phi(x_2)^{p-2-j}_{\sigma_4}\phi(x_1)^{p-2-i}_{\sigma_5},\notag
\end{align}where the first summand of the last line corresponds to the case $\sigma_2=0$ in the second line.

For $II_\sigma$ and $III_\sigma$, we again need \eqref{retpull1}, \eqref{retpull2} and, furthermore, the general commutator identity
\begin{align}\left[\prod\limits_{i=1}^nA_i,\prod\limits_{j=1}^mB_j\right]&=\sum\limits_{k=1}^{n-1}\sum\limits_{l=1}^{m-1}\prod\limits_{i=1}^{k-1}A_i\prod\limits_{j=1}^{l-1}B_j[A_k,B_l]\prod\limits_{j=l+1}^{m}B_j\prod\limits_{i=k+1}^{n}A_i\notag\\
&=\sum\limits_{k=1}^{n-1}\sum\limits_{l=1}^{m-1}\prod\limits_{j=1}^{l-1}B_j\prod\limits_{i=1}^{k-1}A_i[A_k,B_l]\prod\limits_{i=k+1}^{n}A_i\prod\limits_{j=l+1}^{m}B_j
\label{commprod},
\end{align} to obtain

\begin{align}
II_\sigma &= \int\limits_M d_gx_1 \left[\phi(x),\phi(x_1)^{p-1}\right]_{\sigma-1}G_r(x_1,y)\notag\\
&=\int\limits_M d_gx_2 \sum\limits_{\sum\sigma_i=\sigma-1}\sum\limits_{i=0}^{p-2}\phi(x_2)^j_{\sigma_1}\left[\phi(x),\phi(x_2)\right]_{\sigma_2}\phi(x_2)^{p-2-j}_{\sigma_3}G_r(x_2,y)\notag\\
&=\int\limits_M d_gx_2 \sum\limits_{\sum\sigma_i=\sigma-1}\sum\limits_{i=0}^{p-2}\phi(x_2)^j_{\sigma_1}\frac{(-i)^{\sigma_2}}{\sigma_2!}\left\{R_{1,\sigma_2+1}(\phi^\text{in}(x)\,|\,\phi^\text{in}(x_2),{\mathcal L}_\text{int}^{\otimes \sigma_2})\;-\right.\notag\\
&\qquad \left.- \;R_{1,\sigma_2+1}(\phi^\text{in}(x_2)\,|\,\phi^\text{in}(x),{\mathcal L}_\text{int}^{\otimes \sigma_2})\right\}\phi(x_2)^{p-2-j}_{\sigma_3}G_r(x_2,y)\notag\\
&=(p-1)i\int\limits_M d_gx_1 \left\{G_r(x,x_1)-G_a(x,x_1)\right\}G_r(x_1,y)\phi(x_1)^{p-1}_{\sigma-1}\;+\notag\\
&\qquad+\;\int\limits_{M^2} d_gx_1 d_gx_2 \sum\limits_{\sum\sigma_i=\sigma-1}\sum\limits_{i,j=0}^{p-2}\phi(x_1)^i_{\sigma_1}\phi(x_2)^j_{\sigma_2}\frac{(-i)^{\sigma_3}}{\sigma_3!}\;\times\notag\\
&\qquad\qquad\times\;\left\{G_r(x,x_1)R_{1,\sigma_3+1}(\phi^\text{in}(x_1)\,|\,\phi^\text{in}(x_2),{\mathcal L}_\text{int}^{\otimes \sigma_3})\;-\right.\notag\\
&\qquad\qquad \left.- \;G_a(x,x_1)R_{1,\sigma_3+1}(\phi^\text{in}(x_2)\,|\,\phi^\text{in}(x_1),{\mathcal L}_\text{int}^{\otimes \sigma_3})\right\}\phi(x_2)^{p-2-j}_{\sigma_4}\phi(x_1)^{p-2-i}_{\sigma_5}G_r(x_2,y),\notag\end{align}

\begin{align}
III_\sigma &= \int\limits_M d_gx_1 G_a(x,x_1)\left[\phi(x_1)^{p-1},\phi(y)\right]_{\sigma-1}\notag\\
&=\int\limits_M d_gx_1 G_a(x,x_1) \sum\limits_{\sum\sigma_i=\sigma-1}\sum\limits_{i=0}^{p-2}\phi(x_1)^j_{\sigma_1}\left[\phi(x_1),\phi(y)\right]_{\sigma_2}\phi(x_1)^{p-2-j}_{\sigma_3}\notag\\
&=\int\limits_M d_gx_1\,G_a(x,x_1) \sum\limits_{\sum\sigma_i=\sigma-1}\sum\limits_{i=0}^{p-2}\phi(x_1)^j_{\sigma_1}\frac{(-i)^{\sigma_2}}{\sigma_2!}\left\{R_{1,\sigma_2+1}(\phi^\text{in}(x_1)\,|\,\phi^\text{in}(y),{\mathcal L}_\text{int}^{\otimes \sigma_2})\;-\right.\notag\\
&\qquad \left.- \;R_{1,\sigma_2+1}(\phi^\text{in}(y)\,|\,\phi^\text{in}(x_1),{\mathcal L}_\text{int}^{\otimes \sigma_2})\right\}\phi(x_1)^{p-2-j}_{\sigma_3}\notag\\
&=(p-1)i\int\limits_M d_gx_1\, G_a(x,x_1)\left\{G_r(x_1,y)-G_a(x_1,y)\right\}\phi(x_1)^{p-1}_{\sigma-1}\;+\notag\\
&\qquad+\;\int\limits_{M^2} d_gx_1 d_gx_2\,G_a(x,x_1) \sum\limits_{\sum\sigma_i=\sigma-1}\sum\limits_{i,j=0}^{p-2}\phi(x_1)^i_{\sigma_1}\phi(x_2)^j_{\sigma_2}\frac{(-i)^{\sigma_3}}{\sigma_3!}\;\times\notag\\
&\qquad\qquad\times\;\left\{R_{1,\sigma_3+1}(\phi^\text{in}(x_1)\,|\,\phi^\text{in}(x_2),{\mathcal L}_\text{int}^{\otimes \sigma_3})G_r(x_2,y)\;-\right.\notag\\
&\qquad\qquad \left.- \;R_{1,\sigma_3+1}(\phi^\text{in}(x_2)\,|\,\phi^\text{in}(x_1),{\mathcal L}_\text{int}^{\otimes \sigma_3})G_a(x_2,y)\right\}\phi(x_2)^{p-2-j}_{\sigma_4}\phi(x_1)^{p-2-i}_{\sigma_5}.\notag
\end{align}

\begin{figure}
\includegraphics[width=350pt]{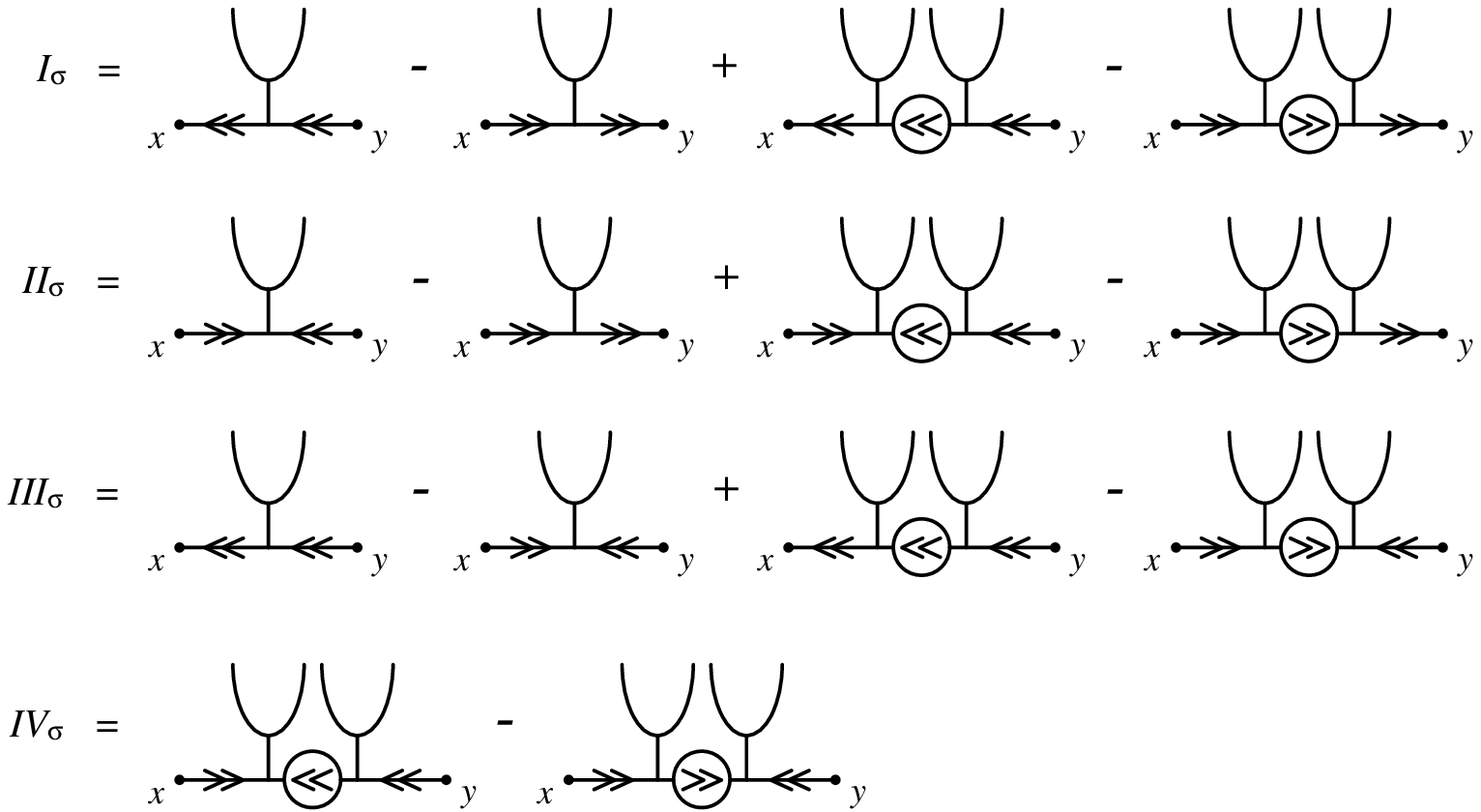}\caption{The partial results for $\sigma>1$}\label{partialccr}\end{figure}

Finally, using \eqref{commprod}, the computation of $IV_\sigma$ yields

\begin{align}
IV_\sigma &= \int\limits_{M^2}d_gx_1d_gx_2\;G_a(x,x_1)\left[\phi(x_1)^{p-1}, \phi(x_2)^{p-2}\right]_{\sigma-2}G_r(x_2,y)\notag\\
&= \int\limits_{M^2}d_gx_1d_gx_2\;G_a(x,x_1)\sum\limits_{\sum\sigma_i=\sigma-2}\sum\limits_{i,j=0}^{p-2}\phi(x_1)^i_{\sigma_1}\phi(x_2)^j_{\sigma_2}\;\times\notag\\
&\qquad\times\;\left[\phi(x_1), \phi(x_2)\right]_{\sigma_3}\phi(x_2)^{p-2-j}_{\sigma_4}\phi(x_1)^{p-2-i}_{\sigma_5}G_r(x_2,y)\notag\\
&=\int\limits_{M^2}d_gx_1d_gx_2\;G_a(x,x_1)\sum\limits_{\sum\sigma_i=\sigma-2}\sum\limits_{i,j=0}^{p-2}\phi(x_1)^i_{\sigma_1}\phi(x_2)^j_{\sigma_2}\;\times\notag\\
&\qquad\times\;\frac{(-i)^{\sigma_3}}{\sigma_3!}\left\{R_{1,\sigma_3+1}(\phi^\text{in}(x_1)\,|\,\phi^\text{in}(x_2),{\mathcal L}_\text{int}^{\otimes\sigma_3})\;-\right.\notag\\
&\qquad\qquad\left.-\;R_{1,\sigma_3+1}(\phi^\text{in}(x_2)\,|\,\phi^\text{in}(x_1),{\mathcal L}_\text{int}^{\otimes\sigma_3})\right\}\phi(x_2)^{p-2-j}_{\sigma_4}\phi(x_1)^{p-2-i}_{\sigma_5}G_r(x_2,y).\notag
\end{align}
We have subsumed the partial results graphically in figure \ref{partialccr}, where the encircled double arrows depict $G_r$-chains. It is straightforward to check the cancellations $I_\sigma-II_\sigma-III_\sigma+IV_\sigma=0$. This closes the proof of the "out"-field CCR.

\paragraph*{Non-quasifree representation}
One of the main claims of this article is that, on non-stationary spacetimes, the outgoing field is in general in the GNS-representation of a state which is {\em not} quasifree. The reason for this is the lack of both spectral conditions and energy-momentum conservation, which would assure the vanishing of higher order truncated Wightman functions of the "out"-field in the stationary case. 

To see this explicitely, let us consider as an example for a non-stationary spacetime ${\mathbb R}^d$
with metric $g(\epsilon)\dfff(1+\epsilon h)\eta$, where $\eta$ is the
Minkowski metric and $h$ a $C^\infty$-function on $M$ of compact support.
One then has $d_gx=\sqrt{\vert\mathsf g\vert}\,dx=(1+\epsilon h)^{d/2} dx$ and such a spacetime is non-stationary since the metric depends on "time".

By means of the methods described in section \ref{calculation_of_wf} (see \cite{Hack} for a detailed calculation), one computes the truncated 4-point function of the "out"-field
 to first order in $-\lambda$ in $\phi^4$ theory on this spacetime as
\begin{equation}
\label{eqa_nontriv} \langle\Omega,\phi^\text{out}(x_1)\phi^\text{out}(x_2)\phi^\text{out}(x_3)\phi^\text{out}(x_4)\Omega\rangle^T_1=12 {\rm Im}\int\limits_{\R^d} dy\,
\prod_{l=1}^4D_{g(\epsilon)}^-(x_l,y)\, (1+\epsilon h(y))^{\frac{d}{2}},
\end{equation}
where the $D^-$ bear the subscript $g(\epsilon)$ to emphasize that they depend on the metric via
the Klein-Gordon-equation. To calculate (\ref{eqa_nontriv}) up to first order in $\epsilon$, one
needs to expand $(1+\epsilon h)^{d/2}$ and $D^-_g$ in $\epsilon$, where $D^-_g(x,y)=\langle\Omega,\phi^{\rm in}_g(y)\phi^{\rm in}_g(x)\Omega\rangle$ can be expanded
in $\epsilon$ by expanding each of the two fields $\phi^{\rm in}_g$ separately via
 the free Klein-Gordon equation. Denoting the expansion of $\phi^{\rm in}_g$ up to first order in
$\epsilon$ as
$$
\phi^{\rm in}_g=\phi^{\rm in}_0+\epsilon\phi^{\rm in}_1+\mathcal{O}(\epsilon^2),$$
we obtain the expansion of $D^-_g$ to first order in $\epsilon$ as
\begin{equation*}
\begin{split}
D^-_g(x,y)
&=\langle\Omega,\phi^{\rm in}_0(y)\phi^{\rm in}_0(x)\Omega\rangle+\epsilon\left(\langle\Omega,\phi^{\rm in}_1(y)\phi^{\rm in}_0(x)\Omega\rangle+\langle\Omega,\phi^{\rm in}_0(y)\phi^{\rm in}_1(x)\Omega\rangle\right)+\mathcal{O}(\epsilon^2)\\
&\fd D^-_{0,0}(x,y)+\epsilon\left(D^-_{0,1}(x,y)+D^-_{1,0}(x,y)\right)+\mathcal{O}(\epsilon^2).
\end{split}
\end{equation*}

To compute the single terms in the expansion of $\phi^{\rm in}_g$, one first
evaluates the wave operator as
\begin{equation*}
\begin{split}
\Box & =-{\vert\mathsf g\vert}^{-\frac{1}{2}}\partial_{b}g^{bc}{\vert\mathsf g\vert}^{\frac{1}{2}}\partial_{c}\\
&=-\eta^{bc}\partial_{b}\partial_{c}+\epsilon\left[-h\eta^{bc}\partial_{b}\partial_{c}-\left(\frac{d}{2}+1\right)\eta^{bc}\left(\partial_{b}h\right)\partial_{c}\right]\\
&\fd\Box_0+\epsilon\Box_1.
\end{split}
\end{equation*}
Inserting this into the Klein-Gordon equation (with minimal coupling) for $\phi^{\rm in}_g$, one obtains to zeroth order in $\epsilon$
$$\left(\Box_0+m^2\right)\phi^{\rm in}_0=0,$$
{\it i.e.}, the Klein-Gordon equation on flat spacetime, and to first order in $\epsilon$
\begin{equation}
\label{eqa_in_field_first_order}
\begin{split}
&\quad\Box_1 \phi^{\rm in}_0 + \left(\Box_0+m^2\right)\phi^{\rm in}_1=0\\
\Rightarrow&\quad\left(\Box_0+m^2\right)\phi^{\rm in}_1=\left[\left(\frac{d}{2}+1\right)\eta^{bc}\left(\partial_{b}h\right)\partial_{c}+h\Box_0\right]\phi^{\rm in}_0\\
\Rightarrow&\quad \phi^{\rm in}_1=G_{r,0}\left[\left(\frac{d}{2}+1\right)\eta^{bc}\left(\partial_{b}h\right)\partial_{c}-hm^2\right]\phi^{\rm in}_0.
\end{split}
\end{equation}
with $G_{r,0}$ denoting the retarded Green's function on Minkowski spacetime.

The well-known expression for $D^-$ on flat spacetime is 
\begin{equation}
D^-_{0,0}(x,y)=\int\limits_{\R^{d-1}}  \frac{d\vec{k}}{2\omega_{\vec{k}}}\,e^{-i\vec{k}\vec{x}+i\omega_{\vec{k}}x_0}\,e^{i\vec{k}\vec{y}-i\omega_{\vec{k}}y_0},
\end{equation}
from which it is explicitly seen that the Fourier transform of $D^-_{0,0}(x,y)$ w.r.t. $x$,
respectively $x-y$, has support in the negative mass shell and its Fourier transform w.r.t.
$y$ has support in the positive mass shell, {\it i.e.}, $D^-_{0,0}(x,y)$ fulfils the spectral condition.
 From (\ref{eqa_in_field_first_order}) it follows that
 $D^-_{1,0}(x,y)$ ($D^-_{0,1}(x,y)$) can be obtained
by application of the operator
$G_{r,0}\left[\left(\frac{d}{2}+1\right)\eta^{ab}\left(\partial_{a}h\right)\partial_{b}-hm^2\right]$
to the first (second) argument of $D^-_{0,0}(x,y)$. Fourier transforming $D^-_{0,1}(x,y)$ on
Minkowski spacetime w.r.t. $y$, one thus gets the Fourier transform of $G_{r,0}$ multiplied by
the Fourier transform of (the derivative of) $h$ convoluted with the Fourier transform of
 (the derivative) of $D^-_{0,0}$. Since the latter convolution smears up the mass shell spectrum
of (the derivative of) $D^-_{0,0}$ and $G_{r,0}$ is known to have off-shell spectrum, the Fourier
transform of $D^-_{0,1}(x,y)$ w.r.t. to $y$ clearly has off-shell support. In contrast,
 the Fourier
transform of $D^-_{1,0}(x,y)$ w.r.t. to $y$ remains on-shell like the
Fourier transform of $D^-_{0,0}(x,y)$.

With these considerations in mind, we can continue examining (\ref{eqa_nontriv}).
An expansion up to first order in $\epsilon$ yields a zeroth order term and
three different first order terms, {\it viz.},
\begin{equation}
\label{eqa_nontriv_expanded}
\begin{split}
& {\rm Im}\int\limits_{\R^d}dy\,
\prod_{l=1}^4D_{g(\epsilon)}^-(x_l,y)\, (1+\epsilon h(y))^{\frac{d}{2}} \\
=\;& {\rm Im}\left[\int\limits_{\R^d}dy\,
\prod_{l=1}^4D_{0,0}^-(x_l,y) + \epsilon\left(\frac{d}{2}\int\limits_{\R^d}
\prod_{l=1}^4D_{0,0}^-(x_l,y)h(y)dy\right.\right.\\
& \qquad\left.\left.+ \sum_j\int\limits_{\R^d}dy\,
D_{0,1}^-(x_j,y)\prod_{l\neq j}D_{0,0}^-(x_l,y)  + \sum_j\int\limits_{\R^d}dy\,
D_{1,0}^-(x_j,y)\prod_{l\neq j}D_{0,0}^-(x_l,y)\right)\right]+\mathcal{O}(\epsilon^2).\\
\end{split}
\end{equation}
\noindent Upon Fourier transforming w.r.t. $y$, it is easily seen that the zeroth order term vanishes due to energy-momentum conservation and the spectral support properties of $D^-_{0,0}$. Similarly, owing to the spectral support properties of  $D_{1,0}^-$, the last first order term in \eqref{eqa_nontriv_expanded} also vanishes due to energy-momentum conservation. However, the remaining two first order terms are in general non-vanishing: regarding $$\int\limits_{\R^d}dy\,
\prod_{l=1}^4D_{0,0}^-(x_l,y)h(y),$$ we can see that it does not vanish in general, despite the spectral properties of $D_{0,0}^-$,  as energy-momentum conservation is violated due to $h$ not having "$\delta$-support" in momentum space. In contrast to this, $$\int\limits_{\R^d}dy\,
D_{1,0}^-(x_j,y)\prod_{l\neq j}D_{0,0}^-(x_l,y)$$ is non-vanishing because of the spectral properties of $D_{1,0}^-$, even if energy-momentum conservation holds. It might be possible to fine-tune the situation in such a way that the two abovementioned non-trivial contributions to $\langle\Omega,\phi^\text{out}(x_1)\phi^\text{out}(x_2)\phi^\text{out}(x_3)\phi^\text{out}(x_4)\Omega\rangle^T_1$ due to alleviated spectral properties on the one hand and abolished energy-momentum conservation on the other hand cancel each other, but in general this is certainly not the case.

To close this section, we remark that from the discussion of the above example it follows that the metric $g$, and hence the curvature of
spacetime, must show characteristic changes on a time scale $t\lessapprox1/(4m)$ to allow the violation
of energy-momentum conservation (and presumably also the deviation from the spectrum condition) to be big enough to assure a non-quasifree "out"-state, {\it e.g.}, $t\lessapprox7\times10^{-6}s$ for a pion with
$m\approx140$MeV. This time scale is sufficiently shorter than the period of nucleosynthesis at about 1 to $10^2$ seconds after the Big Bang, such that these findings are not in contradiction with well established physical facts. It is however significantly longer than the time of, {\it e.g.}, electro-weak symmetry breaking, which happened at $\approx 10^{-12}s$ such that it seems highly  reasonable that full curved spacetime QFT calculations are required to model the physics of the very early universe.  

\section{Unitary transformations between CCR
representations}\label{unitrafo} In the previous section we have seen how non-quasifree states (for free fields) naturally appear in scattering theory on non-stationary curved spacetimes. If one is interested in a scattering picture in terms of particles, one thus needs a way to calculate the particle content of a non-quasifree state, {\it i.e.}, a unitary transformation relating the GNS-representations of a non-quasifree and a quasifree state, but of course the ability to calculate such a transformation is also interesting in its own right. Hence, given a quasifree representation
of the CCR with Fock-space ${\cal F}$, in this section, we
calculate the particle content of non-quasifree representations
that are unitarily equivalent to the given quasifree one. To arrive at such a result, we shall use the language of Wightman
functionals and $\star$-product calculus, {\it cf.}, appendix \ref{star.app} for
a short introduction and notational conventions.

To start, let $\varphi(x)$ be the operator valued distribution fulfilling
the Klein-Gordon equation and the CCR that has been obtained via
the GNS-construction from some Wightman functional (not necessarily corresponding to a quasifree state) 
$\underline{W}'$ with GNS-Hilbert space ${\cal H}_\text{GNS}$ and GNS-vacuum state $\Psi_0\in{\cal H}_\text{GNS}$. Furthermore, let $\xi(x) $ be the
operator valued distribution obtained from a quasifree Wightman
functional $\underline{W}$ via the Fock construction given in
section \ref{setting} with $\Omega\in{\cal F}$ the Fock vacuum.
A relevant application is of course $\varphi=\phi^{\rm
out}$ and $\xi=\phi^\text{in}$. We assume
unitary equivalence of both CCR representations in the following
technical sense: let $U: {\cal F}_{GNS}\to{\cal F}$ be a unitary
transformation such that $U\varphi(f)U^*=\xi(f)$ $\forall f\in {\mathcal D}(M)$ and let $\Psi\dfff U\Psi_0\in{\cal F}$ such that $\Psi$ is in
a dense core of some closure of the Fock creation and annihilation
operators $a(\psi)$ and $a^\dagger(\chi)$, $\psi\in{\cal
H}^+,\chi\in{\cal H}^-$. It is furthermore assumed that, for any
vector $\Upsilon$ in this core, the vectors $a^\sharp(\psi_1)\cdots a^\sharp(\psi_n)\Upsilon $
are jointly continuous in the $\psi_l$
w.r.t. the $({\cal H}^\pm)^{\otimes n}$ and the ${\cal F}$ topologies, where $a^\sharp$ stands for either $a$ or $a^\dagger$.

To determine $U$, it is enough to calculate $\Psi$ since
$U\varphi(f_1)\cdots\varphi(f_n)\Psi_0=\xi(f_1)\cdots\xi(f_n)\Psi$, $f_l\in{\cal D}(M)$,
can be calculated using (\ref{infield.eqa}), once the $n$-particle
components of $\Psi$ are know. Being an element of ${\cal F}$, $\Psi$ can be parameterized as
$$\Psi=\sum_{n=0}^\infty\int\limits_{M^{n}} d_gx_1\cdots d_gx_n\,f_n(x_1,\ldots,x_n) \xi(x_1)\cdots\xi(x_n),$$
where the complex functions $f_n$ are symmetric under permutation
of arguments, purely positive frequency, {\it i.e.}, ${\cal S}_-^{\otimes n}f_n=0$, and fulfil a normalization
condition, {\it viz.}, $\|\Psi\ {\bf 1}_{\cal
F}^2=\sum_{n=0}^\infty\|{\cal S}^{\otimes n}f_n\|^2_{({\cal H}^+)^{\otimes n}}\linebreak =1$.  Furthermore,
the $f_n$ are taken from some function space s.t.
$f_n\mapsto {\cal S}^{\otimes n}f_n\in({\cal H}^+)^{\hat \otimes n}$ is
onto, where $\hat\otimes$ denotes the symmetric tensor product. Given $\ul{W}$ and $\ul{W'}$, computing $U$ is hence equivalent to determining $f_n$, or rather ${\cal S}_+^{\otimes n}f_n$, since the mapping ${\cal S}_+$ is not one-to-one and only the solution part of $f_n$ is "visible" in $\Psi$. 

Let $\ul{f}=(f_0,f_1,f_2,\ldots)$, then obviously
\begin{equation}\ul{W}'=\stackrel{\rightarrow}{D}_{\ul{f}^*}\stackrel{\leftarrow}{D}_{\ul{f}}\ul{W} =\sum_{n,j=0}^{\infty}\stackrel{\rightarrow}{D}_{f_n^*}\stackrel{\leftarrow}{D}_ {f_j}\ul{W},\label{operation}\end{equation}
where the convergence of the infinite sums on the right hand side
follows from the assumption that $\Psi$ is in a core for the {\em
closed} creation and annihilation operators.
The operators $\stackrel{\leftrightarrow}{D}_{\ul{f}}=\sum_{n=0}^\infty\stackrel{\leftrightarrow}{D}_{f_n}$
 act by inserting $f_n$ into the first/last $n$ arguments of a Wightman function, {\it cf.}, appendix \ref{star.app}. Application of
the relation (\ref{star6.eqa}) derived in that appendix and $\star$-multiplication with
$\ul{W}^{\star-1}=\exp_{\star}(-\ul{W}^T)$ yields
\begin{eqnarray}
\label{Fock3.eqa}
\exp_\star(\underline{W}^{'T}-\ul{W}^T)&=&\sum_{m,j=0}^\infty \int\limits_{M^{n+j}}d_gx_1\cdots d_gx_{n+j}\,\sum_{I\in{\cal P}(\{1,\ldots,n+j\}\atop I=\{I_1,\ldots,I_k\},\, k\geq 1}\star_{\,l=1}^{\,k}\stackrel{\leftrightarrow}{D}_{I_l}\ul{W}^T\times\nonumber\\
&&\qquad\times~ f^*_n(x_1,\ldots,x_n)f_j(x_{n+1},\ldots,x_{n+j}).
\end{eqnarray}
Both $\ul{W}^{'}$ and $\ul{W}$ induce CCR representations. By
\cite[Lemma 5.2]{GT}, this is equivalent to $\text{Im}W_2^{'T}=\frac{1}{2}D$ and $W_n^{'T}(x_1,\ldots,x_n)$ being
symmetric under permutation of arguments -- and hence real by the
Hermiticity of $\ul{W}'$ -- for $n\geq3$. This automatically holds
for the quasifree state $\ul{W}$ since $W_2^{T}=D^+$ and
$W_n^T(x_1,\ldots,x_n)=0$ for $n\geq3$. Hence, the left hand side
of (\ref{Fock3.eqa}) is real and symmetric. We note that
\begin{eqnarray}
\label{Fock4.eqa}
\left.
\begin{array}{rcll}
\stackrel{\leftrightarrow}{D}_{I_l} \ul{W}^T&=&0&\mbox{ for } |I_l|>3\\
\stackrel{\leftrightarrow}{D}_{I_l} \ul{W}^T&=&(D^+(x_{j_1},x_{j_2}),0,\ldots)&\mbox{ for } I_l=\{j_1,j_2\}, j_1<j_2\\
\stackrel{\rightarrow}{D}_{I_l} \ul{W}^T&=&(0,D^+(x_{j},\,\cdot\,),0,\ldots)&\mbox{ for } I_l=\{j\}\\
\stackrel{\leftarrow}{D}_{I_l} \ul{W}^T&=&(0,D^+(\,\cdot\,,x_j),0,\ldots)&\mbox{ for } I_l=\{j\}\\
\end{array}
\right\}.
\end{eqnarray}
As a result, only the partitions that consist of sets with one or two elements only contribute in (\ref{Fock3.eqa}). Given such a
partition $I=\{I_1,\ldots,I_l\}$, let $S\dfff\cup_{l:|I_l|=1}I_l$ and
$\hat I\in{\cal P}'(\{1,\ldots,n+j\}\setminus S\})$ the remainder,
which is a pairing partition. Employing this notation, we can can compute
\begin{equation}
\label{Fock6.eqa} \sum_{I\in{\cal P}(\{1,\ldots,n+j\}\atop
I=\{I_1,\ldots,I_k\},\, k\geq
1}\star_{\,l=1}^{\,k}\stackrel{\leftrightarrow}{D}_{I_l}\ul{W}^T=\sum_{\begin{array}{c}
\scriptstyle S\subseteq\{1,\ldots,n+j\}\\\scriptstyle \hat I\in{\cal P'}(\{1,\ldots,n+m\}\setminus S)\\
\scriptstyle\hat I=\{I_1,\ldots,I_{(n+j-|S|)/2}\}\\ \scriptstyle
I_l=\{i_l,j_l\},\,
i_l<j_l\end{array}}\prod_{l=1}^{(n+j-|S|)/2}D^+(x_{i_l},x_{j_l})\,
\star_{j\in S}\stackrel{\leftrightarrow}{D}_{x_j}\ul{W}^T.
\end{equation}
Clearly, $(\star_{j\in
S}\stackrel{\leftrightarrow}{D}_{x_j}\ul{W}^T)_s=0$ if $s\not=|S|$
and for $s=|S|$, $S=\{j_1,\ldots,j_s\}$, $j_1<j_2<\ldots<j_q\leq
n<j_{q+1}<\ldots<j_s$,
\begin{equation}
\label{Fock7.eqa}
\left(\star_{j\in S}\stackrel{\leftrightarrow}{D}_{x_j}\ul{W}^T\right)_s(y_1,\ldots,y_s)=\sum_{\pi\in S_s} \prod_{l=1}^qD^+(x_{j_l},y_{\pi(l)})\prod_{l=q+1}^{s}D^-(x_{j_l},y_{\pi(l)}),
\end{equation}
where $S_n$ denotes the permutations of $\{1,\dots,n\}$. Inserting (\ref{Fock6.eqa}) and (\ref{Fock7.eqa}) into (\ref{Fock4.eqa}) yields
$$\exp_\star\left(\ul{W}^{'T}-\ul{W}^T\right)_s(y_1,\ldots,y_s)=$$
\begin{eqnarray}
&=&\sum\limits_{r\ge s\atop r-s~\text{even}}\sum\limits_{n=0}^r\int\limits_{M^{r}}d_gx_1\cdots d_gx_r \sum\limits_{\begin{array}{c}\scriptstyle \{j_1,\ldots,j_s\}\subseteq \{1,\ldots,r\}\\ \scriptstyle j_1<\cdots<j_q\leq n<\\\scriptstyle~~~ <j_{q+1}<\cdots<j_s\end{array}}\sum\limits_{\begin{array}{c}\scriptstyle \hat I\in{\cal P'}(\{1,\ldots,n+m\}\setminus S)\\
\scriptstyle\hat I=\{I_1,\ldots,I_{(r-s)/2}\}\\ \scriptstyle I_l=\{i_l,k_l\},\, i_l<k_l\end{array}}\sum\limits_{\pi\in S_s}\quad \times\label{Fock8.eqa}\\&&
\times \quad \prod\limits_{l=1}^{(r-s)/2}D^+(x_{i_l},x_{k_l})\, \prod\limits_{l=1}^qD^+(x_{j_l},y_{\pi(l)})\prod\limits_{l=q+1}^{s}D^-(x_{j_l},y_{\pi(l)})\quad \times\nonumber\\&&
\times \quad f_n^*(x_1,\ldots,x_n)f_{r-n}(x_{n+1},\ldots,x_r).\nonumber
\end{eqnarray}
We note that $\int_M d_gx\,D^{\pm}(x,y) f(x)=0$ if $f$ is
positive/negative frequency, {\it cf.}, (\ref{infield.eqa}). Furthermore, by our assumptions, 
$f^*_n$ is purely negative frequency and $f_{n-r}$ purely positive
frequency. One can thus replace all propagators $D^\pm$
in (\ref{Fock8.eqa}) by the real symmetric function $\tilde
D=D^++D^-$ since the integral over the added propagator $D^\mp$
with $f^*_n$ or $f_{n-r}$ always vanishes.

Having done so, we can commute the sums over $n$ and over $S$ on
the right hand side of (\ref{Fock8.eqa}), such that the integral
contains the function
\begin{equation}
\label{Fock9.eqa}
\tilde z_r(x_1,\ldots,x_r)\dfff\sum_{n=0}^rf_n^*(x_1,\ldots,x_n)f_{n-r}(x_{n+1},\ldots,x_r).
\end{equation}
Next, we would like to symmetrise this expression, {\it viz.}, $$z_r(x_1,\ldots,x_r)\dfff(r!)^{-1}
\sum_{\pi\in S_r}\tilde
z_r(x_{\pi(1)},\ldots,x_{\pi(r)}),$$ a procedure which also makes $z_r$ a real
function, and to replace $\tilde z_r$ by $z_r$. To see that this is well-defined, let
$1\leq j<r$. We have to show that $\tilde z_r$ is integrated
w.r.t. a function which is symmetric in $x_j$ and $x_{j+1}$. Given
one term in the combinatorial sum, suppose that $j,j+1\in S$. Then, 
symmetry follows from summation over $S_s$. Next, suppose
that either $j$ or $j+1$ is a member of a pairing and the other index is in
$S$. Then, there exists another contribution to the combinatorial
sum where $j$ and $j+1$ are exchanged showing symmetry for this
case. Let finally $j$ and $j+1$ both be members a pairing. If the
pairings are different, the argument just given applies. If this
is one and the same pairing, then symmetry follows from the
symmetry of $\tilde D$.

Taking into account that the sum over $S_s$ yields 
a factor $s!$, the sum over $S$ a factor $\left({r\atop
s}\right)$ and the sum over pairings a factor
$2^{s-r}(r-s)!/((r-s)/2)!$, one obtains a combinatorial factor
$c_{s,r}$ by multiplication of these contributions. These considerations finally
lead to
\begin{eqnarray}
\label{Fock10.eqa}
\exp_\star\left(\ul{W}^{'T}-\ul{W}^T\right)_s(y_1,\ldots,y_s)&=&\sum_{r=s\atop r-s ~\text{even}}^\infty c_{s,r}\int\limits_{M^{r}}d_gx_1\cdots d_gx_r\,\prod_{l=1}^{(r-s)/2}\tilde D(x_{2l-1},x_{2l}) \;\times\\
&&\qquad\qquad\times\prod_{l=r-s+1}^r\tilde D(x_l,y_{l-r+s}) \,z_r(x_1,\ldots,x_r).\nonumber
\end{eqnarray}

\begin{figure}
\label{triang}
\centerline{\scalebox{.7}{\includegraphics{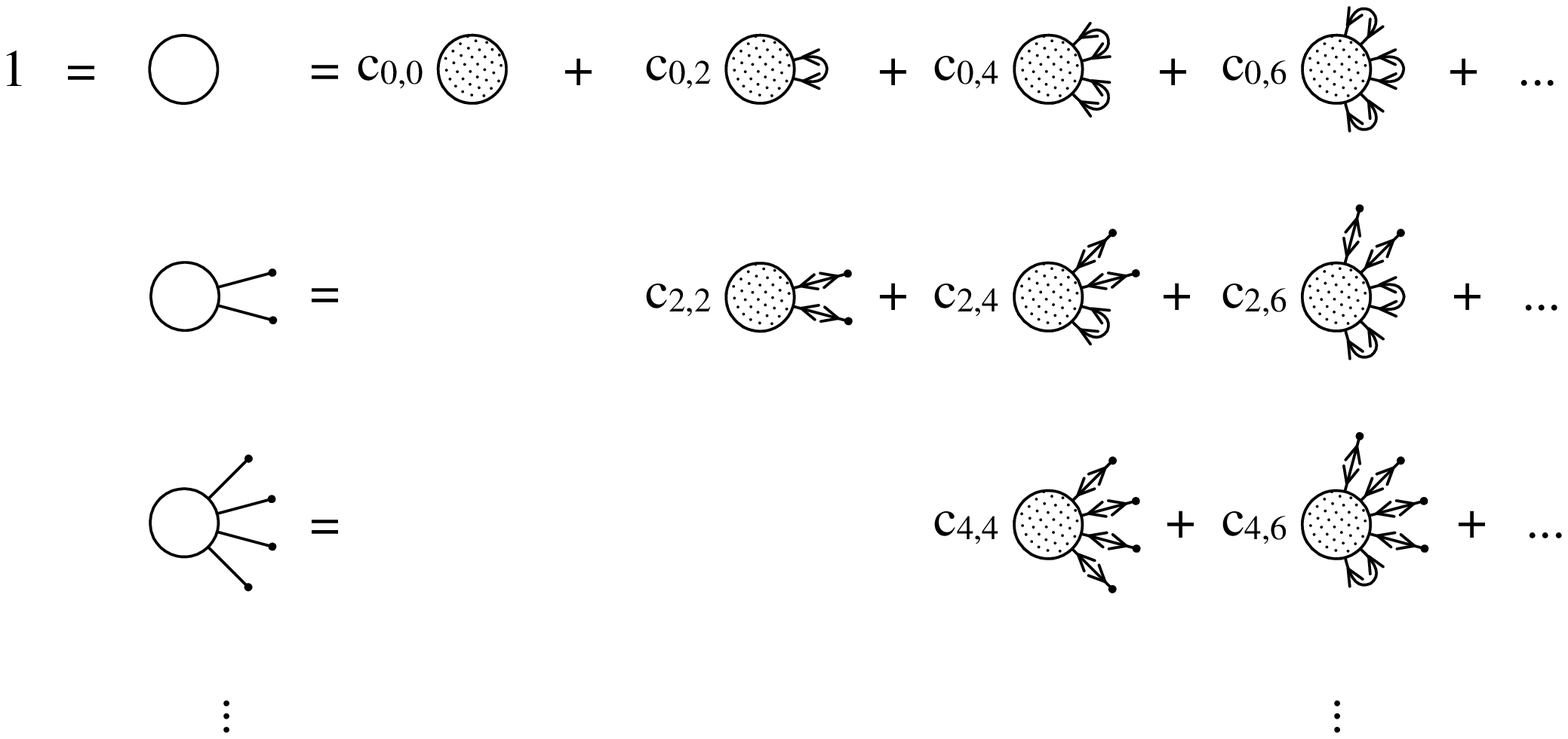}}}
\caption{The triangular system of equations for the functions
${\cal S}^{\otimes r}z_r$ }
\end{figure}

To fulfil our task to compute $U$, we need to solve this system of equations for the solution part of
$z_r$, {\it i.e.}, for ${\cal S}^{\otimes r}z_r=\tilde D^{\otimes r}z_r$. To this end, to obtain a better understanding of the structure of
(\ref{Fock10.eqa}), we introduce some additional graphical notation: we denote the $s$-point function of the
functional on the left hand side by a white circle with $s$ legs
and the function $z_r$ by a shaded circle with $r$ amputated legs.
The integrations with the propagators $\tilde D$ then either add
free legs that carry two arrows with opposite direction or a line
of that type that goes back into the shaded circle. ${\cal S}^{\otimes r}z_r$ thus corresponds to a shaded circle with $r$
legs with double arrows of opposite direction. This way, one obtains two
decoupled systems, one for $s$ even and one for $s$ odd,  {\it cf.}, 
figure \ref{triang} for the even system, which makes the upper
triangular structure visible. In the following, we focus on solving
the even system, the odd system can be solved alike. In
$\phi^p$-theories with $p$ even, the odd system is identically
zero on the left hand side and hence gives ${\cal S}^{\otimes
r}z_r=0$ for odd $r$.
We note that the empty circles are solutions of the Klein-Gordon
equation in each of their legs. By the demand of continuity of the
creation/annihilation operators in $\psi$ and $\chi$ when
repeatedly applied to $\Psi$, Riesz' lemma implies that the
empty circle with $s$ legs is in ${\cal H}^{\hat \otimes s}$. Let
$\{h_j\}_{j\in\N}$ be an ONS in ${\cal H}$. Taking the scalar
product with $h_j$ in the first two legs and then summing over $j$
on the right hand side induces an opposite double arrow line that
goes back into the shaded circle, since $(\tilde Df,\tilde
Dh)=\tilde D(f,h)$ for $f,h$ real. On the left hand side, we
denote this contraction operation by a arrow-less line going back
into the white circle.
\begin{figure}

\centerline{\scalebox{.7}{\includegraphics{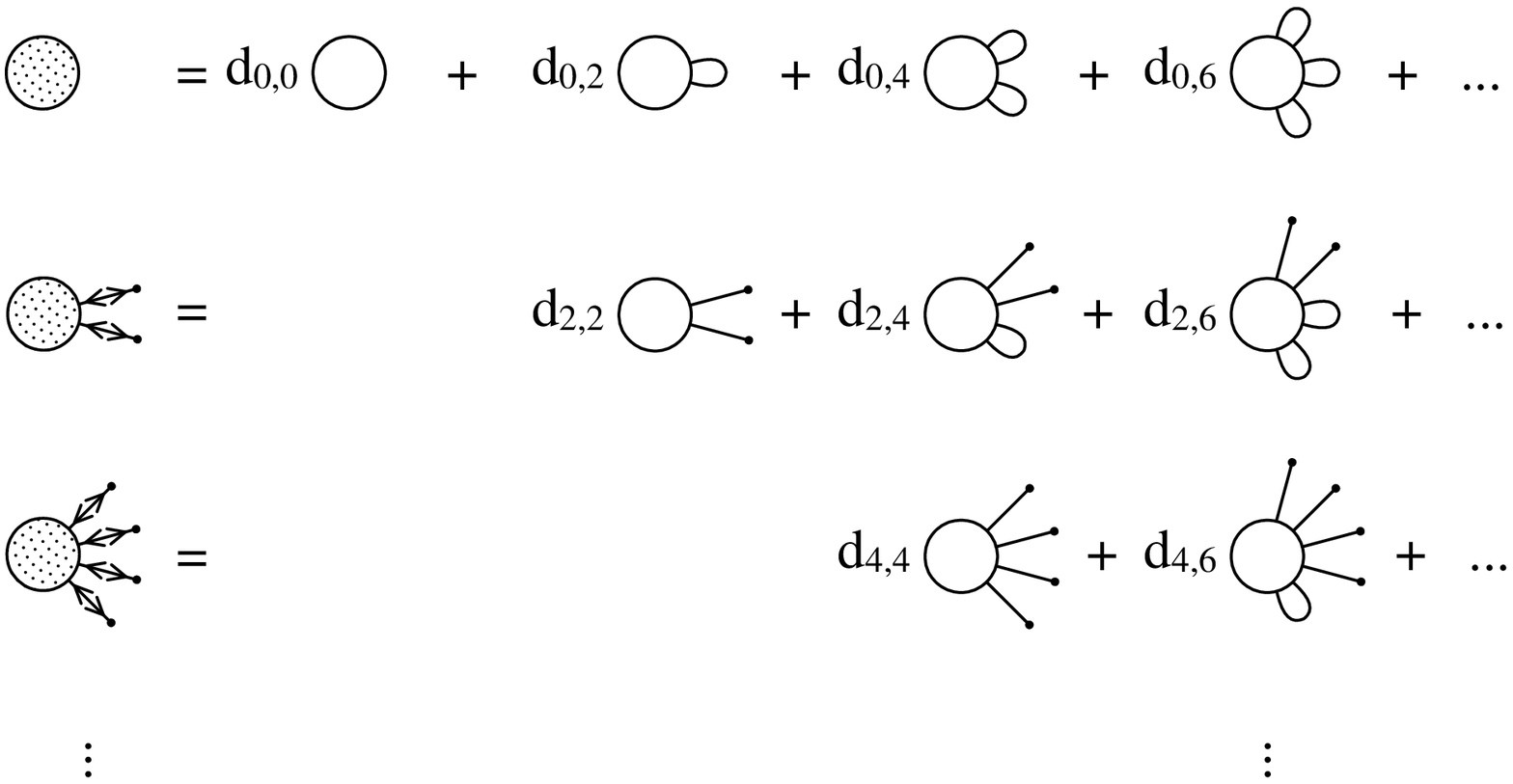}}}
\caption{Solution to the triangular system  }\label{triang2}
\end{figure}

The unique solution of the even system may hence be written down in
graphical form as in figure \ref{triang2}. The solution exists by
assumption of unitary equivalence, thus, all infinite sums involved in the inverse system
converge, which follows from $\lim_{n\to\infty}\Pi(n)\Psi=\Psi$ in
${\cal F}$, where $\Pi(n)$ projects on states with at most $n$
particles, and the fact that for a state with at most $n$ particles
we have a finite system of equations. The constants $d_{s,r}\dfff (C^{-1})_{s,r}$ are
defined as the entries of the inverse of the upper triangular
matrix $C\dfff(c_{s,r})_{s,r\in2\N}$.

Let us recall that the functions $z_r$ have been a convenient intermediate tool, and that our ultimate aim is to determine the solution part of the (purely positive frequency) functions $f_n$. Hence, it remains to reconstruct ${\cal S}^{\otimes n}f_n={\cal S_+}^{\otimes n}f_n$ from the functions ${\cal S}^{\otimes r}z_r$. To achieve this, let us first suppose that
$z_0=|f_0|^2\not=0$. As the state $\Psi$ is only determined up to
a phase, one may assume $f_0>0$. Then, by (\ref{Fock9.eqa}), 
$$
{\cal S}_+^{\otimes r}z_r(x_1,\ldots,x_r)=f_0{\cal S}_+^{\otimes r}f_r(x_1,\ldots,x_r)\,,~~\mbox{ for } r\in\N.
$$
If ${\cal S}^{\otimes r}z_r=0$ for $r<r_0$ and $r_0$ is maximal,
$r_0$ must be even. It follows that ${\cal S}^{\otimes n}f_n=0$
for $n<r_0/2$. Hence, there exist $y_1,\ldots,y_{r_0/2}\in M$ such that
${\cal S}_-^{\otimes\frac{r_0}{2}}\otimes{\cal S}_+^{\otimes
\frac{r_0}{2}}z_{r_0}(y_1,
\ldots,y_{r_0/2},y_1, \ldots,y_{r_0/2})=|{\cal S}^{\otimes
\frac{r_0}{2}} f_{r_0/2}(y_1,\ldots,y_{r_0/2})|^2>0$. We may fix
the phase such that ${\cal S}^{\otimes
\frac{r_0}{2}}f_{r_0/2}(y_1,\ldots, y_{r_0/2})>0$ and we obtain
the solution part of $f_n$, $n\geq r_0/2$, via
$$
{\cal S}_-^{\otimes\frac{r_0}{2}}\otimes{\cal S}_+^{\otimes n}z_{\frac{r_0}{2}+n}(y_1,\ldots,y_{r_0/2},x_1,\ldots,x_n)
={\cal S}_+^{\otimes \frac{r_0}{2}}
f_{r_0/2}(y_1,\ldots,y_{r_0/2}){\cal S}_+^{\otimes
n}f_n(x_1,\ldots,x_n),$$
which closes the looked-for computation of $U$.

To obtain a complete description of the scattering process on non-stationary spacetimes in terms of particles, we have to assume that the spacetime under consideration is asymptotically flat\footnote{Here, asymptotically flat is meant in a rather loose sense, {\it i.e.}, we assume that both in the remote future and in the remote past of $(M,g)$, there is an open, non-empty, globally hyperbolic subset of (M,g) which contains a Cauchy surface of $(M,g)$ and is isometric to such a subset of Minkowski spacetime. In this setting, it is straightforward to define preferred states as the pull-backs of the ones in Minkowski space. However, even within the more strict definition of asymptotically flat spacetimes, one can obtain preferred states, as devised in \cite{Dappiaggi}.} in the remote future and past, such that unique preferred quasifree states are available both for the incoming and the outgoing field. Then, there are associated Fock spaces, say ${\cal F}_\text{in}={\cal F}$ and ${\cal F}_\text{out}$, and a combination of the scattering theory described in the previous sections and the results obtained in this section gives the $n$-particle amplitudes $f_n$ of the scattered incoming quasifree state in the particle picture of the remote future. If one wants to determine particle production from an incoming multi-particle state, one can apply suitably smeared incoming fields $\phi^\text{in}(x)$ to the incoming vacuum, then calculate the outgoing representation of the CCR, and then conclude as above.

\section{Conclusions and outlook}
\label{conc}

In the this work, we have seen that non-quasifree states for free fields appear naturally in scattering theory on non-stationary curved spacetimes. This result is well in line with recent works \cite{HR, Sanders} which show that a certain class of non-quasifree states, namely, the ones for which the truncated 2-point function is a distribution with the singularity structure of the Minkowski vacuum state and the other truncated $n$-point functions are smooth, is the natural class of states in perturbative quantum field theory on curved spacetimes. In the light of this, it seems somewhat unnatural and unnecessary to restrict oneself to quasifree states, although some important technical results are only available for quasifree states, see, {\it e.g.}, \cite{Lueders, Ver}.

Therefore, and also because there are situations where one is interested in the particle interpretation of non-quasifree states, we develop a method to calculate, provided it exists, a unitary transformation relating a non-quasifree state to a quasifree one. Heuristically, the form of the result could have been anticipated: as we assume unitary equivalence, the GNS-vacuum associated to the non-quasifree state corresponds to a state in the Fock space related to the quasifree state under consideration, and the task is to compute the $n$-particle components $f_n$ of this state. Since we assume both states to fulfil the same commutation relations, they only differ in the real and symmetric part of their 2-point function and the higher order truncated $n$-point functions, which are real and symmetric in the non-quasifree case and vanishing in the quasifree case. It is thus not surprising that our result  \eqref{Fock10.eqa} relates the truncated $n$-point functions of the non-quasifree state to an expression in the symmetric part of the 2-point function of the quasifree state smeared with a real and symmetrised version of the $f_n$. The non-trivial part of our result are, however, the combinatorics involved, and we have managed to tame them by encoding them conveniently into $\star$-calculus on the dual of the Borchers-Uhlmann algebra.

The method of computing a unitary transformation relating the GNS-representations of non-quasifree and quasifree states developed in this work is well-suited for general treatments of the topic, but not for explicit numerical calculations. A different method to compute such a transformation, which is based on \cite{Glauber}, works for finite-dimensional systems, {\it i.e.}, "mode-by-mode", and is therefore better suited for numerical computations, has been developed and applied in \cite{Hack}.

\begin{acknowledgments}For the first named author it is
a pleasure to thank Horst Thaler for the ongoing and very fruitful
exchange of ideas. T.H. would like to express his gratitude towards Nicola Pinamonti for the interesting and fruitful discussions regarding graph combinatorics. We also thank Klaus Fredenhagen for a very instructive discussion on asymptotic conditions.
\end{acknowledgments}

\appendix

\section{$\star$ -calculus}
\label{star.app}
In this appendix, we present some useful applications of $\star$-calculus and the related notation we have used in the main body of this work.  $\star$-calculus goes back to Borchers \cite{Bo} and Ruelle \cite{Ru}; for $\star$-products in algebraic quantum field theory, {\it cf.}, \cite{Brunetti} and the references cited therein, for $\star$-calculus for quantum fields in the context of Hopf algebras see, {\it e.g.}, Mestre and Oeckl \cite{MO}.

Let $\underline{\cal D}$ be the Borchers-Uhlmann algebra of the Hermitian scalar field with multiplication $\otimes$ and unit $ {\bf 1}\dfff(1,0,\ldots)$, {\it i.e.}, the non-commutative, unital, involutive, topological, free tensor algebra over the space of complex valued test functions ${\cal D}(M)$.
If $\underline{f}\in\underline{\cal D}$ is a monomial, $\underline{f}=(0,\ldots,0,f_n,0,\ldots)$ then we identify $\underline{f}$ with $f_n\in{\cal D}(M^{n})$.  We also note that the involution $*$ acts via $f_n^*(x_1,\ldots,x_n)=\overline{f_n(x_n,\ldots,x_1)}$. For $f_j\in {\cal D}(M)$ and $N\subset \N$ finite, we define $\otimes_{j\in N}f_j$ ($\otimes_{\emptyset}f_j\dfff {\bf 1}$) such that
the tensor product preserves the natural order of $N$.  A co-commutative co-product $\Delta:\ul{\cal D}\to\ul{\cal D}\otimes\ul{\cal D}$ can then be defined by
\begin{equation}
\label{star1.eqa}
\Delta(\otimes_{n\in N}f_n)=\sum_{S\subseteq N}\left(\otimes_{n\in S}f_n\right)\otimes\left(\otimes_{n\in N\setminus S}f_n\right),
\end{equation}
linearity and continuity. Note that, in (\ref{star1.eqa}), the tensor products in the parantheses are multiplication in $\underline{\cal D}$, whereas the tensor product
between the paratheses is the one in $\underline{\cal D}\otimes\underline{\cal D}$. Furthermore, the projection $\varepsilon((f_0,f_1,\ldots))=f_0$ defines a co-unit.

Let $\underline{\cal D}'=\{(W_0,W_1,\ldots): W_0\in\C,W_n\in{\cal D}'(M^{\times n}), n\geq 1\}$ be the topological dual of $\underline{\cal D}$, {\it i.e.}, the space of Wightman functionals. In many applications, a Wightman functional will be given by the sequence of $n$-point (truncated) VEVs, {\it viz.}, $W^{(T)}_n(x_1,\ldots,x_n)\dfff\langle\Omega,\varphi(x_1)\cdots\varphi(x_n)\Omega\rangle^{(T)}$ for some operator valued distribution $\varphi$.  The co-product on $\ul{\cal D}$ naturally leads to a product, $\star$, which can be defined as $$\underline{W}\star\underline{V}\dfff(\underline{W}\otimes\underline{V})\circ \Delta,$$ making $\underline{\cal D}'$ a commutative algebra with unit $ {\bf 1}=\varepsilon$.

From (\ref{star1.eqa}) we get
$$(\underline{W}\star\underline{V})_{| N|}(\otimes_{n\in N} f_n)=\sum_{S\subseteq N}W_{ |S|}(\otimes_{n\in S}f_n)V_{|N|-| S|}(\otimes_{n\in N\setminus S}f_n)$$
which shows the coincidence of $\star$ with Borchers' $s$-product.

It is easy to see that $(\underline{W}^{\star n})_m=0$ for $n>m$ and $\underline{W}\in \underline{\cal D}_1'\dfff\{\underline{V}\in\underline{\cal D}'\,\vert\,\overline V_0=0\}$. Hence, arbitrary $\star$-power series converge on $\underline{\cal D}_1'$. In particular, the $\star$-exponential $\exp_\star:\underline{\cal D}_1'\to {\bf 1}\oplus\underline{\cal D}_1'$  and $\star$-logarithm $\log_\star: {\bf 1}\oplus\underline{\cal D}_1'\to\underline{\cal D}_1'$
are well defined through their power series $\exp_\star(W)\dfff\sum_{n=0}^\infty W^{\star n}/n!$ and $\log(W)\dfff-\sum_{n=1}^\infty ( {\bf 1}-W)^{\star n}/n$. Furthermore, exp$_\star$ and log$_\star$ are inverses of one another and the usual relations hold, 
{\it i.e.}, $\exp_\star(\underline{W}+\underline{V})=\exp_\star\underline{W}\star\exp_\star\underline{V}$, $\exp_\star(0)= {\bf 1}$ and $\log_\star(\underline{W}\star\underline{V})=\log_\star\underline{W}+\log_\star\underline{V}$. For $\log_\star \underline{W}$, we also use the notation $\underline{W}^T$. In fact,
if $\underline{W}$ denotes the collection of Wightman functions, $(\underline W^T)_n$ is the truncated $n$-point function defined in equation (\ref{def_truncated_wf}) above, {\it cf.}, \cite{Bo}.

Let $f\in{\cal D}(M)$, and $\stackrel{\rightarrow}{m}(f),\stackrel{\leftarrow}{m}(f):\ul{\cal D}\to\ul{\cal D}$ be the left and right multiplication with $f$ in $\ul{\cal D}$. We then define left and right derivatives $\stackrel{\rightarrow}{D}_f$ and $\stackrel{\leftarrow}{D}_f$ on $\ul{\cal D}'$ via  $\stackrel{\rightarrow}{D}_f\ul{W}\dfff\ul{W}\circ \stackrel{\rightarrow}{m}(f)$ and $\stackrel{\leftarrow}{D}_f\ul{W}\dfff W\circ \stackrel{\leftarrow}{m}(f)$. In fact, it is easily verified that a Leibniz rule holds for these derivatives (which of course motivates this nomenclature), {\it viz.},
$$
\stackrel{\leftrightarrow}{D}_f(\ul{W}\star\ul{V})=(\stackrel{\leftrightarrow}{D}_f\ul{W})\star \ul{V}+\ul{W}\star(\stackrel{\leftrightarrow}{D}_f\ul{V})
$$
where $\stackrel{\leftrightarrow}{D}_f$ stands either for $\stackrel{\rightarrow}{D}_f$ or for $\stackrel{\leftarrow}{D}_f$. For a formal power series $h(t)$, this implies that $\stackrel{\leftrightarrow}{D}_fh_\star(\ul{W})=(h')_\star(\ul{W})\star \stackrel{\leftrightarrow}{D}_f\ul{W}$, where $f_\star$ denotes the formal $\star$-power series one obtains from the formal power series $f$ by replacing the normal product with the $\star$-product and $f'(t)\dfff f^{(1)}(t)\dfff df/dt$. For notational convenience, we shall sometimes write $\stackrel{\leftrightarrow}{D}_f\fd\int_Md_gx\,f(x) \stackrel{\leftrightarrow}{D}_x$.

Higher order derivative operators $\stackrel{\leftrightarrow}{D}_{\ul{f}}\ul{W}\dfff\ul{W}\circ\stackrel{\leftrightarrow}{m}(\ul{f})$, $\ul{f}\in\ul{\cal D}$, can also be written as $\stackrel{\leftrightarrow}{D}_{\ul{f}}=\sum_{n=0}^\infty \stackrel{\leftrightarrow}{D}_{f_n}$, and we write 
$$\stackrel{\leftrightarrow}{D}_{f_n}\fd\int\limits_{M^{n}}d_gx_1\cdots d_gx_n\,f(x_1,\ldots x_n)\stackrel{\leftrightarrow}{D}_{x_1}\cdots\stackrel{\leftrightarrow}{D}_{x_n}.$$
By induction, it is easy to see that, for $f_n$ and $f_j$ symmetric, the following chain rule holds
\begin{eqnarray*}
\stackrel{\rightarrow}{D}_{f_n}\stackrel{\leftarrow}{D}_{f_j}h_\star(\ul{W})&=&\sum_{k=1}^{n+j}(h^{(k)})_\star(\ul{W})\star\int\limits_{M^{n+j}}d_gx_1\cdots d_gx_{n+j}\,\sum_{I\in{\cal P}(\{1,\ldots,n+j\})\atop I=\{I_1,\ldots,I_k\}}\star_{\,l=1}^{\,k}\stackrel{\leftrightarrow}{D}_{I_l}\ul{W}\times\\
&&\qquad\times~ f_n(x_1,\ldots,x_n)f_j(x_{n+1},\ldots,x_{n+j}),
\end{eqnarray*}
where ${\cal P}(N)$ is the set of partitions of $N$ and $\stackrel{\leftrightarrow}{D}_{I_l}\dfff\stackrel{\rightarrow}{D}_{x_{j_1}}\cdots \stackrel{\rightarrow}{D}_{x_{j_q}}\stackrel{\leftarrow}{D}_{x_{s_1}}\cdots \stackrel{\leftarrow}{D}_{x_{s_r}}$ for $I_l=\{j_1,\ldots,j_q,s_1,\ldots,s_r\}$, $1\leq j_1<j_2<\ldots<j_q\leq n<s_1<s_2<\ldots<s_r\leq n+j$.

For $h(t)=\exp(t)$, $\underline{W}^T\in\ul{\cal D}_1'$, and $\ul{W}=\exp_\star(\ul{W}^T)$, we finally obtain
\begin{eqnarray}
\label{star6.eqa}
\stackrel{\rightarrow}{D}_{f_n}\stackrel{\leftarrow}{D}_{f_j}\ul{W}&=&\ul{W}\star\int\limits_{M^{\times(n+j)}}d_gx_1\cdots d_gx_{n+j}\,\sum_{I\in{\cal P}(\{1,\ldots,n+j\}\atop I=\{I_1,\ldots,I_k\}, k\geq 1}\star_{\,l=1}^{\,k}\stackrel{\leftrightarrow}{D}_{I_l}\ul{W}^T\times\nonumber\\
&&\qquad\times~ f_n(x_1,\ldots,x_n)f_j(x_{n+1},\ldots,x_{n+j}),
\end{eqnarray}
which closes this appendix.

\section{Identities involving retarded products}
\label{retapp}

In this appendix, we would like to prove some identities we have used both implicitely and explicitely in the proof of the CCR of the outgoing field in the main body of
this work.

First, we need to show that the $\sigma$-th order of the $j$-th power of the interacting field is, like in the case of the
interacting field itself, the result of the $\sigma$-fold retarded action of ${\mathcal L}_\text{int}$ on the $j$-th power of the "in"-field. For this purpose, it is convenient to introduce a shorthand notation for the multiple application of the formal  differential operator ${\mathcal D}(x)$, {\it viz.},
$${\mathcal D}^nA(x)\dfff \int\limits_{M^n} d_gX_{n,1}\,{\mathcal D}^n(X_{n,1})A(x){\bf 1}_{X_{n,1}\succeq x}\dfff\int\limits_{M^n}\prod\limits_{i=1}^nd_gx_i\,{\mathcal D}(x_n)\cdots{\mathcal D}(x_1)A(x) {\bf 1}_{x_n\succeq\cdots\succeq x_1\succeq x},$$ where $X_{k,l}$ denotes the causally ordered $k-l+1$-tuple $(x_k,x_{k-1},\dots,x_l) {\bf 1}_{x_k\succeq\cdots\succeq x_l}$ and $A(x)$ is an arbitrary expression in terms of "in"-fields. Later, we will also need tuples of spacetime points devoid of mutual causal relations. To this end, we will denote by $X^{\displaystyle{\not}\succeq}_{k,l}$ the $k-l+1$-tuple $(x_k,x_{k-1},\dots,x_l)$. With this notation, we have to show that
$$\phi(x)^j_\sigma\dfff\sum\limits_{\sum\sigma_i=\sigma}\prod\limits_{i=1}^j\phi(x)_{\sigma_i}={\mathcal D}^\sigma \phi^\text{in}(x)^j.$$
On can achieve this by an induction similar to the one employed to show that $\phi(x)_\sigma$ defined via the Yang-Feldman equation is equal to ${\mathcal D}^\sigma \phi^\text{in}(x)$. The case $\sigma=0$ is clear. Let us assume the result holds for $\sigma$ and show how its validity for $\sigma+1$ follows. We can compute \begin{align}{\mathcal D}^{\sigma+1}\phi^\text{in}(x)^j&={\mathcal D}{\mathcal D}^{\sigma}\phi^\text{in}(x)^j\notag\\
&=\int\limits_{M}d_gx_{\sigma+1}\,{\mathcal D}(x_{\sigma+1})\sum\limits_{\sum\sigma_i=\sigma}\prod\limits_{i=1}^j\left(\int\limits_{M^{\sigma_i}}d_g
X_{\sigma_i,1}\,{\mathcal D}^{\sigma_i}(X_{\sigma_i,1})\phi^\text{in}(x) {\bf 1}_{x_{\sigma+1}\succeq X_{\sigma_i,1}\succeq x}\right)\notag\\
&=\int\limits_{M}d_gx_{\sigma+1}\,\sum\limits_{\sum\sigma_i=\sigma}\sum\limits_{k=1}^j\prod\limits_{i=1}^{k-1}\left(\int\limits_{M^{\sigma_i}}d_g
X_{\sigma_i,1}\,{\mathcal D}^{\sigma_i}(X_{\sigma_i,1})\phi^\text{in}(x) {\bf 1}_{x_{\sigma+1}\succeq X_{\sigma_i,1}\succeq x}\right)\times\notag\\
&\qquad\qquad\times\;\int\limits_{M^{\sigma_k}}d_g
X_{\sigma_k,1}\,{\mathcal D}^{\sigma_k+1}(x_{\sigma+1}\times X_{\sigma_k,1})\phi^\text{in}(x) {\bf 1}_{x_{\sigma+1}\succeq X_{\sigma_k,1}\succeq x}\times\label{indmon1}\\
&\qquad\qquad\times\;\prod\limits_{i=k+1}^j\left(\int\limits_{M^{\sigma_i}}d_g
X_{\sigma_i,1}\,{\mathcal D}^{\sigma_i}(X_{\sigma_i,1})\phi^\text{in}(x) {\bf 1}_{x_{\sigma+1}\succeq X_{\sigma_i,1}\succeq x}\right).\notag\end{align}
In comparison, we know that $\phi(x)^j_{\sigma+1}$ equals
\begin{align}
\sum\limits_{\sum\sigma_i=\sigma+1}\prod\limits_{i=1}^j\left(\int\limits_{M^{\sigma_i}}d_g
X_{\sigma_i,1}\,{\mathcal D}^{\sigma_i}(X_{\sigma_i,1})\phi^\text{in}(x) {\bf 1}_{X_{\sigma_i,1}\succeq x}\right)\label{indmon2},
\end{align}
notably, the appearing $j$ integrands in every summand of $\sum_{\sum\sigma_i=\sigma+1}$ are mutually independent, while the corresponding integrands in \eqref{indmon1} depend on one another, namely, only one of them depends on $x_{\sigma+1}$ explicitely, but all others are constrained such that their integration domains are causally later than $x_{\sigma+1}$. If we can show that all combinatorially possible constraints appear exactly once, then their sum yields the independent integrands of \eqref{indmon2}. To achieve this, let us pick an arbitrary but fixed summand of $\sum_{\sum\sigma_i=\sigma+1}$, say, $(\sigma_1,\dots,\sigma_j)=(n_1,\dots,n_j)$. Apart from integration domain restrictions, the $j$ integrands of this summand are the same as the ones of all summands of $\sum_{\sum\sigma_i=\sigma}\sum_{k=1}^j$ in \eqref{indmon1} with $(\sigma_1,\dots,\sigma_k+1,\dots,\sigma_j)=(n_1,\dots,n_j)$. If we assume that $m$ entries of $(n_1,\dots,n_j)$ are non-zero, then there are exactly $m$ summands of the latter type due to the exhaustion of the sum $\sum_{\sum\sigma_i=\sigma}$ and the Leibniz rule for ${\mathcal D}(x_{\sigma+1})$. The integrands of these $m$ summands are such that it is always a different of the $j$ integrands that explicitely depends on $x_{\sigma+1}$. Therefore, all possibilities of the event "One of the $j$ integrands involves an integration variable which is causally earlier than the integration domains of all other $j-1$ integrands." appear, and their sum yields exactly the summand of \eqref{indmon2} under consideration; this implies ${\mathcal D}^{\sigma+1}\phi^\text{in}(x)^j=\phi(x)^j_{\sigma+1}$.

In the following, we can use this result to prove the looked-for recursive relations

$$\frac{(-i)^n}{n!}R_{1,n+1}\left(\phi^\text{in}(x)\,|\,\phi^\text{in}(y), {\mathcal L}_\text{int}^{\otimes n}\right)$$
\begin{equation}\label{retpulla}=\int\limits_M d_gx_1\;G_r(x,x_1)\sum\limits_{\sum \sigma_i = n-1}\sum\limits_{j=0}^{p-2}\phi(x_1)^j_{\sigma_1}\,\frac{(-i)^{\sigma_2}}{\sigma_2!}R_{1,\sigma_2+1}\left(\phi^\text{in}(x_1)\,|\,\phi^\text{in}(y), {\mathcal L}_\text{int}^{\otimes \sigma_2}\right)\,\phi^{p-2-j}_{\sigma_3}(x_1)\end{equation}
\begin{equation}\label{retpullb}=\int\limits_M d_gx_1\sum\limits_{\sum \sigma_i = n-1}\sum\limits_{j=0}^{p-2}\phi(x_1)^j_{\sigma_1}\,\frac{(-i)^{\sigma_2}}{\sigma_2!}R_{1,\sigma_2+1}\left(\phi^\text{in}(x)\,|\,\phi^\text{in}(x_1), {\mathcal L}_\text{int}^{\otimes \sigma_2}\right)\,\phi(x_1)^{p-2-j}_{\sigma_3}\;G_r(x_1,y).\end{equation} To this effect, let us define another formal differential operator $\delta(x)$ via $$\delta(y)A(x)\dfff D(\,\cdot\,, y)\frac{dA(x)}{d\phi^\text{in}(\,\cdot\,)}\dfff i\left[\phi^\text{in}(y),A(x)\right],$$ where $A(x)$ is again an arbitrary expression in terms of incoming fields and the empty argument $(\,\cdot\,)$ denotes that the argument of the "in"-field the differential operators currently acts upon in a Leibniz rule-summand has to be inserted in that slot.

We can now proceed to prove \eqref{retpulla} by induction. In case $n=1$, one can straightforwardly compute that both expressions equal $$(p-1)i\int\limits_Md_gx_1\;G_r(x,x_1)G_r(x_1,y)\phi(x_1)^{p-2}.$$
Assuming that the wished-for equality holds for $n$, we can analyse the case $n+1$, {\it viz.},
$$\frac{(-i)^{n+1}}{n!}R_{1,n+2}\left(\phi^\text{in}(x)\,|\,\phi^\text{in}(y), {\mathcal L}_\text{int}^{\otimes n+1}\right)=$$
\begin{align*}=&\;i\int\limits_{M^{n+1}}\prod\limits_{i=1}^{n+1}d_gx_i\;\sum\limits_{k=1}^{n+1}\left({\mathcal D}^{n+1-k}(X_{n+1,k+1})\delta(y){\mathcal D}^k(X_{k,1})\phi^\text{in}(x){\bf 1}_{X_{n+1,k+1}\succeq y \succeq X_{k,1}\succeq x}\right)\\
=&\;i\int\limits_{M^{n+1}}\prod\limits_{i=1}^{n+1}d_gx_i\;{\mathcal D}(x_{n+1})\sum\limits_{k=1}^{n}\left({\mathcal D}^{n-k}(X_{n,k+1})\delta(y){\mathcal D}^k(X_{k,1})\phi^\text{in}(x) {\bf 1}_{x_{n+1}\succeq X_{n,k+1}\succeq y \succeq X_{k,1}\succeq x}\right)\;+\\
&\qquad + \; i\delta(y)\int\limits_{M^{n+1}}d_gX_{n+1,1}\;{\mathcal D}^{n+1}(X_{n+1,1})\phi^\text{in}(x) {\bf 1}_{y\succeq X_{n+1,1}\succeq x}\\
=&\;i\int\limits_{M^{n+1}}\prod\limits_{i=1}^{n+1}d_gy_i\;{\mathcal D}(x_{n+1})\frac{(-i)^{n+1}}{n!}R_{1,n+1}\left(\phi^\text{in}(x)\,|\,\phi^\text{in}(y), {\mathcal L}_\text{int}^{\otimes n}(X^{\displaystyle{\not}\succeq}_{n,1})\right) {\bf 1}_{x_{n+1}\succeq X^{\displaystyle{\not}\succeq}_{n,1}\times y \succeq x}\;+\\
&\qquad + \; i\delta(y)\int\limits_{M^{n+1}}d_gX_{n+1,1}\;G_r(x,x_1){\mathcal D}^{n}(X_{n+1,2})\phi^\text{in}(x_1)^{p-1} {\bf 1}_{y\succeq X_{n+1,2}\succeq x_1\succeq x}.
\end{align*}The remaining steps are cumbersome but elementary. To write them down at this point, we would have to introduce even more abbreviating notation which is necessary to avoid loosing the overview, we thus prefer to rather sketch them briefly: the next step would be to insert the induction hypothesis in the first term of the last line of the above equation and to rewrite the resulting retarded product of lower order by means of the formal differential operators ${\mathcal D}$ and $\delta$. Employing the Leibniz rule for ${\mathcal D}$ and the combinatorial considerations regarding different possibilities to distribute causal relations among the factors of a product we have already used at the beginning of this appendix, one can show that the two terms in the last line of the above equation indeed add up to
$$\int\limits_M d_gx_1\;G_r(x,x_1)\sum\limits_{\sum \sigma_i = n}\sum\limits_{j=0}^{p-2}\phi(x_1)^j_{\sigma_1}\,\frac{(-i)^{\sigma_2}}{\sigma_2!}R_{1,\sigma_2+1}\left(\phi^\text{in}(x_1)\,|\,\phi^\text{in}(y), {\mathcal L}_\text{int}^{\otimes \sigma_2}\right)\,\phi(x_1)^{p-2-j}_{\sigma_3},$$ which proves the first of the two looked-for recursion relations for $\frac{(-i)^n}{n!}R_{1,n+1}\left(\phi^\text{in}(x)\,|\,\phi^\text{in}(y), {\mathcal L}_\text{int}^{\otimes n}\right)$.

To prove the second one \eqref{retpullb}, we can again perform an induction by order. The case $n=1$ can be validated straighforwardly and for $n+1$ a computation yields like in the preceding proof
$$\frac{(-i)^{n+1}}{n!}R_{1,n+2}\left(\phi^\text{in}(x)\,|\,\phi^\text{in}(y), {\mathcal L}_\text{int}^{\otimes n+1}\right)=$$
\begin{align*}=&\;i\int\limits_{M^{n+1}}\prod\limits_{i=1}^{n+1}d_gx_i\;{\mathcal D}(x_{n+1})\sum\limits_{k=1}^{n}\left({\mathcal D}^{n-k}(X_{n,k+1})\delta(y){\mathcal D}^k(X_{k,1})\phi^\text{in}(x) {\bf 1}_{x_{n+1}\succeq X_{n,k+1}\succeq y \succeq X_{k,1}\succeq x}\right)\;+\\
&\qquad + \; i\delta(y)\int\limits_{M^{n+1}}d_gX_{n+1,1}\;{\mathcal D}^{n+1}(X_{n+1,1})\phi^\text{in}(x) {\bf 1}_{y\succeq X_{n+1,1}\succeq x}.
\end{align*} Employing the by now multiply used combinatorial considerations, one can show that the first summand in the last line equals $$\int\limits_M d_gx_1\sum\limits_{\sum \sigma_i = n}\sum\limits_{j=0}^{p-2}\phi(x_1)^j_{\sigma_1}\,\frac{(-i)^{\sigma_2}}{\sigma_2!}R_{1,\sigma_2+1}\left(\phi^\text{in}(x)\,|\,\phi^\text{in}(x_1), {\mathcal L}_\text{int}^{\otimes \sigma_2}\right)\,\phi(x_1)^{p-2-j}_{\sigma_3}\;G_r(x_1,y)$$ up to some missing combinatorial possibilities. To show that the second summand in the last line accounts for exactly these possibilities, one has to be able to "extract" a $G_r(\,\cdot\,,y)$ from it. This can be achieved by the Leibniz rule for $\delta$. To wit, every ${\mathcal D}$ in $\delta(y){\mathcal D}^{n+1}(X_{n+1,1})\phi^\text{in}(x)$ involves a commutator with ${\mathcal L}_\text{int}=(\phi^\text{in})^p/p$, such that the application of $\delta(y)$ to ${\mathcal D}^{n+1}(X_{n+1,1})\phi^\text{in}(x)$ yields a sum over terms without $\delta(y)$ but with one ${\mathcal D}$ replaced by a commutator with $\delta(y){\mathcal L}_\text{int}$, which equals $G_r(\,\cdot\,,y)\phi^\text{in}(\,\cdot\,)^{p-1}$ due to the enforced causal relations of the arguments. Further elementary steps then lead to the wished-for conclusion.

\vspace{1cm}

\noindent {\sc Hanno Gottschalk}\\
\rm Institut f\"ur angewandte Mathematik\\
Rheinische Friedrich-Wilhelms-Universit\"at Bonn\\
Wegelerstr. 6\\
D-51373 Bonn, Germany\\
e-mail: gottscha@wiener.iam.uni-bonn.de

\vspace{1cm}

\noindent {\sc Thomas-Paul Hack}\\
\rm II. Institut f\"ur Theoretische Physik\\
Universit\"at Hamburg\\
Luruper Chaussee 149\\
22761 Hamburg, Germany\\
e-mail: thomas-paul.hack@desy.de

\end{document}